\documentclass[aps,prx,showpacs,floatfix,superscriptaddress,byrevtex,twocolumn]{revtex4-1}
\pdfoutput=1
\usepackage{soul}
\usepackage{amsmath}
\usepackage{amssymb}
\usepackage{amstext}
\usepackage{amsopn}
\usepackage{amsfonts}
\usepackage{amsxtra}
\usepackage{dcolumn}
\usepackage{graphicx}
\usepackage{bm}
\usepackage{hyperref}
\usepackage{xcolor}
\definecolor{linkcol}{rgb}{0,0,0.4} 
\definecolor{violet}{rgb}{0.32,0,0.52} 
\definecolor{citecol}{rgb}{0.5,0,0} 
\usepackage[expansion=true]{microtype}
\hypersetup{
bookmarksopen=true,
pdftitle="Temperature-dependent ellipsometry measurements of partial Coulomb energy in Superconducting Materials",
pdfauthor="J. Levallois et al.", 
pdfsubject="Temperature-dependent ellipsometry measurements of partial Coulomb energy in Superconducting Materials", 
pdfstartview={FitH},
pdfmenubar=true, 
pdfhighlight=/O, 
colorlinks=true, 
pdfpagemode=UseNone, 
pdfpagelayout=SinglePage, 
pdffitwindow=true, 
linkcolor=blue, 
citecolor=violet, 
urlcolor=violet}
\begin{document}
\title{Temperature-dependent ellipsometry measurements of partial Coulomb energy in superconducting cuprates}
\author{J. Levallois}
\affiliation{Department of Quantum Matter Physics, University of Geneva, Quai Ernest-Ansermet 24, CH-1211 Gen\`{e}ve 4, Switzerland}
\author{M. K. Tran}
\affiliation{Department of Quantum Matter Physics, University of Geneva, Quai Ernest-Ansermet 24, CH-1211 Gen\`{e}ve 4, Switzerland}
\author{D. Pouliot}
\affiliation{Department of Physics, University of Illinois at Urbana-Champaign, 1110 West Green Street, Urbana, Illinois, USA}
\author{C. N. Presura}
\affiliation{Philips Research, Professor Holstlaan 4, 5656 AE Eindhoven, The Netherlands}
\author{L. H. Greene}
\affiliation{Department of Physics, University of Illinois at Urbana-Champaign, 1110 West Green Street, Urbana, Illinois, USA}
\author{J. N. Eckstein}
\affiliation{Department of Physics, University of Illinois at Urbana-Champaign, 1110 West Green Street, Urbana, Illinois, USA}
\author{J. Uccelli}\author{E. Giannini}
\affiliation{Department of Quantum Matter Physics, University of Geneva, Quai Ernest-Ansermet 24, CH-1211 Gen\`{e}ve 4, Switzerland}
\author{G. D. Gu}
\affiliation{Condensed Matter Physics and Materials Science Department, Brookhaven National Laboratory, Upton, NY 11973 5000, USA}
\author{A. J. Leggett}\email{aleggett@illinois.edu}
\affiliation{Department of Physics, University of Illinois at Urbana-Champaign, 1110 West Green Street, Urbana, Illinois, USA}
\author{D. van der Marel}\email{dirk.vandermarel@unige.ch}
\affiliation{Department of Quantum Matter Physics, University of Geneva, Quai Ernest-Ansermet 24, CH-1211 Gen\`{e}ve 4, Switzerland}
\date{\today}
\begin{abstract}
We performed an experimental study of the temperature and doping dependence of the energy-loss function of the bilayer and trilayer Bi-cuprate family. 
The primary aim is to obtain information on the energy stored in the Coulomb interaction between the conduction electrons, on the temperature dependence thereof, and on the change of Coulomb interaction when Cooper-pairs are formed. 
We performed temperature-dependent ellipsometry measurements on several Bi$_2$Sr$_2$CaCu$_2$O$_{8-x}$ single crystals: under-doped with $T_c=60, 70$ and 83~K, optimally doped with $T_c=91$~K, overdoped with $T_c=84, 81, 70$ and $58$~K, as well as optimally doped Bi$_2$Sr$_2$Ca$_2$Cu$_3$O$_{10+x}$ with $T_c=110$~K. 
Our first observation is that, as the temperature drops through $T_c$, the loss function in the range up to 2~eV displays a change of temperature dependence as compared to the temperature dependence in the normal state. 
This effect at -- or close to -- $T_c$ depends strongly on doping, with a sign-change for weak overdoping. The size of the observed change in Coulomb energy, using an extrapolation with reasonable assumptions about its $q$-dependence, is about the same size as the condensation energy that has been measured in these compounds. Our results therefore lend support to the notion that the Coulomb energy is an important factor for stabilizing the superconducting phase. Due to the restriction to small momentum, our observations do not exclude a possible significant contribution to the condensation energy of the Coulomb energy associated to the region of $q$ around $(\pi,\pi)$. 
\end{abstract}
\maketitle
\section{Introduction}
Ever since the discovery of high $T_c$ superconductivity in the cuprates, a large body of theoretical and experimental research has been concentrated on the mechanism of superconductivity.
The primary thermodynamic quantity of interest is the statistical average of the Hamiltonian, $E$.
An isolated system ({\em i.e.} a system in which the entropy is conserved) becomes superconducting if, and only if, $E$ in the superconducting state is more favorable than $E$ of all alternative states of matter.
Starting at the most basic level, the appropriate Hamiltonian for a system of electrons and nuclei consists of two terms, the kinetic energy (of nuclei and electrons) and Coulomb interaction energy (between nuclei and nuclei, nuclei and electrons and electrons and electrons).
At this basic level it follows directly from the virial theorem~\cite{Chester56} that the transition must involve a saving of the Coulomb energy; what is less obvious~\cite{Hirsch92} is whether this is still true when one goes to the more phenomenological level of description standard in solid-state physics, where the relevant ``Coulomb energy'' is only that of the interaction between the conduction electrons. 
Several years ago one of us~\cite{Leggett1999a,Leggett1999b,Leggett1998} postulated that it is indeed the saving of the inter-conduction electron energy, and specifically the part associated with long wavelengths and mid infrared frequencies, which is the main driver of the superconducting transition in the cuprates (the ``MIR scenario''). 
Here we employ a basic result from linear response theory, that the {\em partial} Coulomb energy associated with a given wave vector $\mathbf{q}$ is proportional to a thermally weighted integral of the electron energy loss function $L_\mathbf{q}(\omega)$ over all frequencies, 
\begin{eqnarray}
E_C^{\mathbf{q}}&=&\frac{\hbar}{2\pi}\int_{0}^{\infty} L_\mathbf{q}(\omega)(1+2n_{\omega})
d\omega
\label{E^C_q}
\end{eqnarray}
where $n_{\omega}=1/(\exp(\hbar\omega/k_BT)-1)$. For $q \sim 0$ the relevant loss function is that measured in optical ellipsometry.
We present experimental loss-function spectra measured in this way for a series of high-$T_c$ cuprates with different carrier concentrations; from these data we calculate the partial Coulomb energy $E_C^{0}(T)$ between 15 and 300 Kelvin in 1 Kelvin steps. The temperature dependence of $E_C^{0}$ and $\gamma_C=T^{-1}dE_C^{0}/dT$ reveal the evolution as a function of doping of the changes of Coulomb energy associated with pairing and with the superconducting phase transition. 
The spectroscopic ellipsometry rig used in the present study has the advantage of high stability, high throughput and dense sampling as a function of temperature. 
As a result the energy loss spectra for $q\sim 0$ presented here, and in particular the observed subtle temperature dependencies, provide an important bench-mark for future studies of the Coulomb energy using alternative methods such as transmission electron energy loss spectroscopy (EELS). 

If the original MIR scenario is correct, then one would prima facie expect it to be reflected in a decrease, at and below the superconducting transition, of the loss function in the MIR region of the spectrum as measured in the optics.
Thus the first question (question (A)) which we shall address in this paper is a qualitative one, namely: in the various regions of the phase diagram explored, does the small-$q$ MIR loss function increase, decrease or remain constant (relative to the extrapolated normal-state behavior, see below) at and below $T_c$? 
This question can be answered directly from the experimental data.
Should the answer to this question for some particular value of doping turn out to be that it increases or remains constant, then the prima facie implication (though see below) would seem to be that the MIR scenario cannot explain the mechanism of superconductivity at least in this region of the phase diagram.

Our second question (question (B)) which is prima facie relevant only if (where) a decrease in the loss function is observed, is: is the decrease in the loss function which we measure quantitatively consistent with the MIR scenario, that is, the hypothesis that all or most of the superconducting condensation energy comes from the saving of Coulomb energy in the ``small-$q$'' regime and the MIR frequency region? 
It should be strongly emphasized that an answer to this question requires not only a careful definition of the scenario (in particular what we mean by ``small-$q$''), but also a crucial {\em assumption}, namely that the value of the loss function measured in our optical experiments, for which the ``effective'' $q$ is of the order of the inverse of the high-frequency penetration depth, $\sim 0.002$~\AA$^{-1}$, can be extrapolated to the much larger values of $q$ (up to $\sim 0.31$~\AA$^{-1}$) which dominate the theoretical expression for the Coulomb energy in the MIR scenario.
In the normal phase a comparison of the values of the loss function as measured in optical experiments with that measured in EELS is consistent with such an extrapolation~\cite{nakai1990,presura2003,DvdM2015}. Whether this remains valid for the {\em changes} observed at and below the superconducting transition is a question that needs to be addressed by future EELS experiments. It is worth mentioning in this context that, on the basis of inelastic neutron scattering data of the cuprates~\cite{demler1998}, indications have been obtained for a significant contribution to the condensation energy {from $\mathbf{q} \sim (\pi,\pi)$.}

At this point it may be useful to review briefly the original motivation for the scenario. As we will see in more detail below (see Eq.~\eqref{gq}), the expectation value of the total Coulomb energy can be rigorously expressed as a sum (integral) of contributions from different Fourier components $E_C^{\mathbf{q}}$. 
As explained in Refs.~\onlinecite{Leggett1999a,Leggett1999b}, the starting observation is that one possible origin of the well-known dependence of the superconducting transition temperature $T_c$ on the number of CuO$_2$ layers per unit cell is the effect of the Coulomb interaction between the conduction electrons in different planes.
If indeed the saving of this energy is a major contribution to the increase of $T_c$ (and thus by inference to that of the condensation energy per plane), then since the relevant matrix element falls off as a function of the in-plane wave vector $q$ as $\exp{(-qd)}$ where $d=3.2$~\AA~is the interplane spacing { (within the bilayers)}, it follows that a major contribution to the saving must come from wave vectors $q<q_0=d^{-1} =0.31$~\AA$^{-1}$. 
It is then highly plausible (though, of course, not a rigorous statement) that the same must be true also for the Coulomb energy, {\em i.e.} that a major contribution to the condensation energy comes from { the intraplane Coulomb energy with} $q<q_0$, and we use this { condition on $q$} as one of the defining ingredients in the ``MIR scenario.''

\begin{figure}[h!!!!]
\centerline{
\includegraphics[width=\columnwidth]{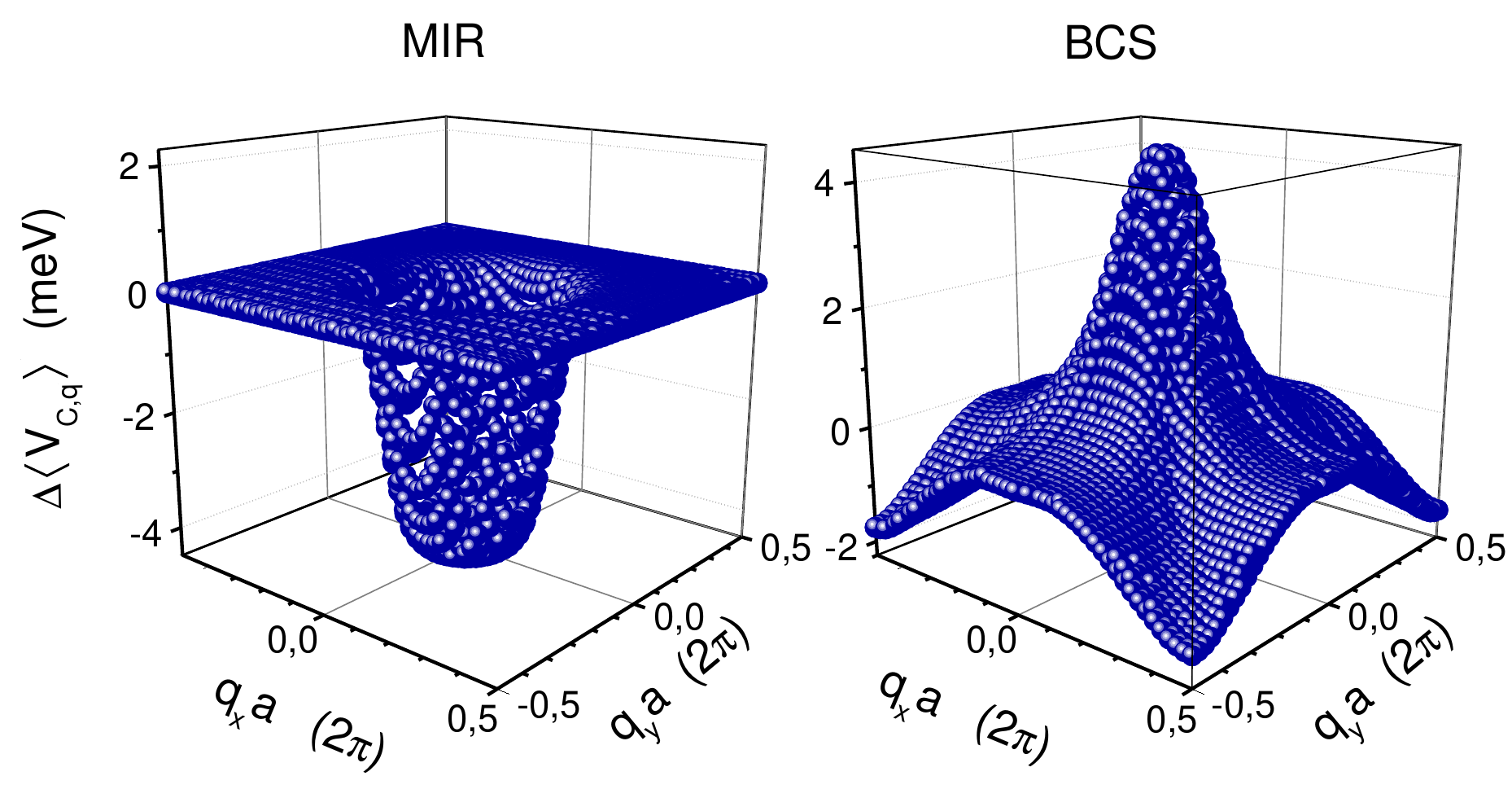}
}
\caption{S--N difference of the partial Coulomb energy $E^{C}_{\mathbf{q}}$. Left: According to the MIR scenario (schematic).
Right: Resulting from a BCS model calculation~\cite{dirk2016} for $d$-wave symmetry, $p=0.16$ hole doping, and the interaction adjusted such as to give $T_c= 100$ K.  
Both panels represent the $q_z=0$ cut in momentum space, corresponding to electric field polarized along the planes.
$E^{C}_{\mathbf{q}}$ at $\mathbf{q}=0$ corresponds to the integral (Eq.~\eqref{E^C_q}) of the optical in-plane loss-function.}
\label{fig:d-wave_S_q}
\end{figure}

In Fig.~\ref{fig:d-wave_S_q} the partial Coulomb energy difference $\Delta E_C^{\mathbf{q}}$ is displayed, where $\Delta$ signifies the value of the superconducting phase minus the one of the normal phase (hereafter ``S--N'' difference for brevity). The result anticipated from the MIR scenario (left) is compared with the behavior expected from BCS theory and later extensions thereof addressing the collective response~\cite{pwa1958,leggett1965,Kostyrko1992,DvdM1995,lee2015} (right). The BCS result~\cite{dirk2016} is a smooth function of $\mathbf{q}$ with a maximum at $q=0$. This example also demonstrates that in the case of $d$-wave pairing BCS theory predicts a negative sign for $\Delta E_C^{\mathbf{q}}$ near $(\pi,\pi)$, a state of affairs that is held responsible for stabilizing the superconducting state in the context of $tJ$ and Hubbard models for high $T_c$~\cite{scalapino1998,demler1998,maier2004,haule2007,gull2012,fratino2015}. Note that at the end of the day this also represents a form of Coulomb energy, which is unfortunately not accessible with optical spectroscopy. 
The attentive reader may object that when the numbers are put in, the BCS value shows substantial variation with $q$ on the scale of the coherence length, {\em i.e.} between the optical regime and $0.31$~\AA$^{-1}$, whereas we are assuming the absence of such variation. There is no contradiction here: in BCS theory the relevant frequencies which contribute are of the order of (the Fermi velocity $v_F$ times) $q$ itself, whereas in the MIR scenario they are substantially larger than $v_Fq$ even for $q\sim 0.3$~\AA$^{-1}$. There is no reason why the $q$-dependence should be the same in these two very different cases.

A second question relates to the region of frequency $\omega$ in which the saving occurs: it was argued in Refs.~\onlinecite{Leggett1999b}, following the ``Willie Sutton principle'', that since the only frequency regime in which the loss function in the normal state is both substantial and likely to have contributions principally from the conduction electrons in the CuO$_2$ planes is the MIR; this is the region where the maximal saving should occur. 
Again, the lower and upper frequency cutoffs are somewhat arbitrary, but a natural definition of the relevant ``MIR frequency regime'' might be say $0.6-1.8$~eV. In summary, the ``canonical'' definition of the original MIR hypothesis is that a very substantial contribution (let us say $>70-80\%$ of the whole) to the total condensation energy is made by a saving of the Coulomb energy associated with wave vectors $q<0.31$~\AA$^{-1}$ and with frequencies $0.6-1.8$ eV, and this constitutes the conjecture addressed by our question (B).
We should further note that in the original formulation the saving was assumed to occur only at and below the macroscopic transition temperature $T_c$.

The above discussion obviously raises a number of further questions which go beyond the original scenario. 
First, what if we relax the constraint $q<q_0$, {\em i.e.} consider the saving of inter-conduction Coulomb energy from all $q$ in the first Brillouin zone: is it enough to constitute the whole of the condensation energy? 
This is an interesting question, but in order to obtain any information on it from the optical data we would need to extrapolate the optically measured dielectric constant to the whole of the zone, which seems implausible, so we will not discus it further here. 
A second generalization would be to raise the same question with the original constraint on $q$ but with the frequency regime extended to lower and/or higher frequencies (perhaps right up to the X-ray regime).
This is our question~(C). 
Finally, we could try to relax the constraint on the relevant temperature regime and consider a generalized scenario~\cite{AJL&DP2} in which all or a substantial part of the energy saving takes place above the macroscopic transition; this is question~(D).
In the following we will attempt to give a definitive answer to question (A), a relatively definitive one (subject to the extrapolation assumption) to question (B), and some information which, while it does not answer questions~(C) or~(D) unambiguously, may be qualitatively relevant to them.
\section{Coulomb energy in superconductors}\label{section2}
\subsection{Sum rules}
To motivate this subsection we briefly recapitulate the fundamental concepts underlying the MIR scenario.
The theoretical description of the conduction electrons in the cuprates is based on the following key assumptions~\cite{Leggett1999a}:
(i)~Core and conduction electrons can be treated as separate systems.
(ii)~The loss spectra below 2~eV are dominated by the $CuO_2$ planes.
(iii)~Ionic motion (phonons) is irrelevant.
(iv)~The optical response and the mechanism of pairing are essentially two-dimensional, {\em i.e.} it is justified to neglect inter multilayer tunneling in the analysis of superconductivity (details are provided in Appendix~\ref{C-axis_contribution}). Thus the generic Hamiltonian is written as
\begin{equation} 
\hat{H}=\hat{T}_{}+\hat{U}+\hat{V}_{C},
\label{hamiltonian}
\end{equation}
with $\hat{T}_{}$ the in-plane kinetic energy, $\hat{U}$ the potential felt by the electrons due to the ionic cores, and $\hat{V}_{C}$ the conduction electron-electron Coulomb interaction energy
~\footnote{This interaction should be taken to be screened by the core electrons.}. 
The main postulate of the MIR scenario is, that the interaction energy $\langle\hat{V}_{C}\rangle$ decreases upon entering the superconducting state.

The first purpose of the present paper is to explore the qualitative consistency of the optical data with the MIR scenario, {\em i.e.} to answer question~(A).
For this limited purpose we ignore complications associated both with the layered nature of the cuprates ({\em i.e.} we treat them for electrodynamic purposes as 3D continua) and with the screening of the Coulomb interaction by the ionic cores (for these complications see~\cite{presura2003} and appendices~\ref{Bound} and~\ref{C-axis_contribution}). 
The total energy {\em per unit cell} contained in the inter-particle Coulomb energy is provided by the relation
\begin{eqnarray}
E_{C}=
\langle\hat{V}_{C}\rangle
&=&
\frac{1}{2N}\sum_{\mathbf{q}} V_\mathbf{q} S_\mathbf{q}
\label{Ec}
\end{eqnarray}
where $N$ is the number of unit cells, $S_\mathbf{q}=\langle \hat{\rho}_{-\mathbf{q}}\hat{\rho}_\mathbf{q}\rangle$ the structure factor and $V_\mathbf{q}=4\pi e^2/ q^2$ the Fourier transform of the Coulomb potential.
For the expression on the right-hand side we can employ the general relation~\cite{NozieresPines1999} between the structure factor and the charge-susceptibility following from the fluctuation-dissipation theorem
\begin{equation}
S_\mathbf{q}=\frac{1}{\pi}\int_{0}^\infty \chi^{\prime\prime}(\mathbf{q}, \omega)
(1+2n_{\omega})
\hbar d\omega
\label{gq}
\end{equation}
The susceptibility appearing in this expression, $\chi (\mathbf{q}, \omega)$, measures the charge response to a density perturbation with frequency $\omega$ and wavevector $\mathbf{q}$, and is related to the longitudinal dielectric function, $\epsilon_{\parallel}(\mathbf{q}, \omega)$, through
\begin{equation}
\frac{1}{\epsilon_{\parallel}(\mathbf{q},\omega)}={1-V_q\chi(\mathbf{q},\omega)}
\label{Eps}
\end{equation}
The imaginary part of Eq.~\eqref{Eps} 
\begin{equation} 
L_\mathbf{q}(\omega)=\text{Im} \frac{-1}{\epsilon_{\parallel}(\mathbf{q},\omega)}
\label{lossfunction}
\end{equation}
can be measured with the help of inelastic electron scattering~\cite{nakai1990}, and is for this reason called the electron energy loss function. 
The corresponding transverse quantity, in which $\epsilon_{\parallel}$ is replaced by $\epsilon_{\perp}$, can be measured in the $q\rightarrow 0$ limit by optical spectroscopy. For normal metals it is well established that in the limit of $q\to 0$ one has $\varepsilon_{\parallel}(\mathbf{q},\omega)=\varepsilon_{\perp}(\mathbf{q},\omega)$. This has recently been proven also for the superconducting state by two of us~\cite{AJL&DP1} in the relevant limit $q\rightarrow 0,~\omega \neq 0$. The equivalence in the case of the cuprates is for example illustrated in Fig.~3 of Ref.~\onlinecite{DvdM2015}. 
Together, Eqs.~\eqref{Ec},~\eqref{gq},~\eqref{Eps}, and~\eqref{lossfunction} provide the ``Coulomb energy sumrule''~\cite{Nozieres1959,NozieresPines1999,Leggett2006,Leggett1999a,Leggett1999b,Leggett1998}, which generalized to finite temperature provides the Coulomb interaction energy
\begin{equation}\label{coulombsumrule}
E_{C}=\frac{1}{ N}\sum_{\mathbf{q}}E_C^{\mathbf{q}}
\end{equation}
where $E_C^{\mathbf{q}}$ is given by Eq.~\eqref{E^C_q}.

It should be noted that neither the form of the Hamiltonian~\eqref{hamiltonian} nor the above derivation of the Coulomb energy sum rule necessarily implies that the standard textbook description of the many-body conduction-electron wave function as approximately a Slater determinant of Bloch waves is a good one. 
However, we can always use the Bloch waves as a basis, and if we do so then one consequence of the occurrence of the periodic crystalline potential $U$ in the Hamiltonian~\eqref{hamiltonian} is the occurrence of ``Umklapp'' scattering processes. 
In a recent study Lee calculated the influence of Umklapp processes on the spectral weight of the loss function near the plasma resonance, and predicted an increase of plasmon spectral weight as the system undergoes the superconducting phase transition~\cite{lee2015}.

For the special limit in which in the band picture ({\em i.e.}~the set of energy eigenstates of the single-particle terms $\hat{T}_{}+\hat{U}$ in Eq.~\eqref{hamiltonian}) the lowest relevant band reduces to a nearest-neighbor tight-binding model, and one assumes that the interacting conduction electrons are confined to this band, there exists a second well-known sum rule for the ``kinetic'' energy per unit cell~\cite{maldague1977,baeriswyl1986,Scalapino1993,Bergeron2011}:
\begin{equation}\label{K-sumrule}
¥{K}=-\frac{\hbar^2 \Omega_0}{\pi^2 e^2 a^2}\int_0^\infty \omega\text{Im}\epsilon(\omega)d\omega 
\end{equation}
where $a$ is the in-plane lattice parameter, and $\Omega_0$ is the unit cell volume.
${K}$ subsumes the contributions from the first two terms of Eq.~\eqref{hamiltonian} in addition to Hartree-Fock (and higher order) contributions from the interaction term. Eq.~\eqref{K-sumrule} is therefore to some extent complementary to the Coulomb-energy sumrule, Eq.~\eqref{coulombsumrule}. Eq.~\eqref{K-sumrule} has been subject of intensive investigations pertaining to the question whether superconductivity in the cuprates is caused by a lowering of ``kinetic'' ({\em i.e.} single-particle) energy~\cite{molegraaf2002,pwa95,Hirsch92}. 
Interestingly, it turns out that in underdoped samples of the cuprates the ``kinetic'' energy behaves oppositely to the BCS prediction ({\em i.e.}~is decreased by the N--S transition), while on the overdoped side it behaves consistently with BCS ({\em i.e.} is increased)~\cite{deutscher2005,carbone2006b}, which is in fact consistent with numerical calculations based on the Hubbard model and the $t-J$ model~\cite{gull2012,fratino2015,haule2007}.
However, one should beware of assuming that the ``kinetic'' (single-particle) energy which enters the sum rule~\eqref{K-sumrule} is necessarily the expectation value of the sum of the single-particle terms $\hat{T}_{}$ and $\hat{U}$ in Eq.~\eqref{hamiltonian}; the tight-binding description leading to Eq.~\eqref{K-sumrule} is at a different level from that of the Hamiltonian~\eqref{hamiltonian}, and it is for example not excluded that the Coulomb term in~\eqref{hamiltonian} may affect the effective tunnelling matrix elements in the tight-binding description.
Thus, should it for example be found experimentally that in some doping interval both the RHS of Eq.~\eqref{coulombsumrule} and the RHS of Eq.~\eqref{K-sumrule} decrease at the N--S transition, this would not necessarily constitute a paradox.
\subsection{Optical data and the MIR scenario}
As mentioned the present work aims at exploring the energy stored in the inter-electronic Coulomb interactions, using precise measurements and analysis of the optical loss function. 
As emphasized in the Introduction, to infer anything about the Coulomb energy from the optical data we need to extrapolate our results to finite $q$. 
The big advantage of optics is the possibility to acquire data during extended periods of time. This allows to obtain detailed information about the relative changes of Coulomb energy as a function of temperature and doping. These results in turn provide a benchmark for the accuracy needed to detect these trends with momentum-sensitive techniques such as inelastic neutron scattering, (resonant) inelastic {X-ray} scattering, or electron energy loss spectroscopy, which are subject to the severe constraints on measurement time inherent to large facilities.

\section{Methods}\label{Methods}
\subsection{Samples}
We investigated high purity single crystals of Bi$_2$Sr$_2$Ca$_2$Cu$_3$O$_{10+x}$ (Bi2223) and of Bi$_2$Sr$_2$CaCu$_2$O$_{8-x}$ (Bi2212), with $ab$-plane oriented surfaces of several mm$^2$. The Bi2212 samples are easily cleavable, providing clean and mirror-like surfaces for optical studies. 
Details on growth and characterization of the crystals are provided in Appendix~\ref{appsample}. 
We use the empirical Tallon-Presland relation between carrier concentration and $T_c$~\cite{tallon1995,presland1991} to determine the doping $p$. For our under and optimally doped Bi2212 samples with $T_c=60, 70, 83$ and $91$~K; this yields $p=0.1, 0.11, 0.13$ and $0.16$, respectively. 
For the overdoped Bi2212 samples with $T_c=84, 81, 70$ and $58$~K, we get $p=0.19, 0.2, 0.21$ and $0.23$, respectively.
\subsection{Ellipsometry measurements}
Using ellipsometry at an angle of incidence of 70$^\circ$ with the surface normal we measured the real and imaginary part of the ratio of $p$-polarized over $s$-polarized complex reflectivity coefficients, $\rho=r_p/r_s$ (see Appendix~\ref{ellipso1}). 
The spectrum is measured continuously while the temperature is varied from 15~K to 300~K at a rate of 0.2~K/min. 
For an isotropic material the dielectric function $\epsilon(\omega, T)$ can be readily obtained by numerical evaluation of the relation
\begin{eqnarray}
\epsilon(\omega,T)&=&\sin^2\theta+\frac{\sin^4\theta}{\cos^2\theta}\left(\frac{1-\rho}{1+\rho} \right)^2 
\label{dielectric}
\end{eqnarray}
\noindent following from the Fresnel equations, where $\theta$ is the angle of incidence, 
and the loss function  is 
\begin{eqnarray}
L(\omega,T)&=&\mbox{Im} \frac{-1}{\epsilon(\omega,T)}
\label{lossf}
\end{eqnarray}
For an optically anisotropic material the expressions are more complicated, but often, as in the present case, the {$c$-axis} admixture is small, and can be corrected easily using a rapidly converging iterative method which is outlined in Appendix~\ref{ellipso2}. 
The loss function yields, through frequency integration, the partial Coulomb energy, Eq.~\eqref{E^C_q}.
For brevity we will drop the momentum $\mathbf{q}$ in the subsequent discussion of the optical properties, with the understanding that the experimental data presented here are representative of $q\sim0$.
For the purposes of the present study it is convenient to define also an integral between (sample-dependent) limits $\omega_1, \omega_2$ which are chosen so that the integration runs over a frequency interval where the loss function has the same sign of temperature variation (in our case, it increases upon cooling); {\em i.e.} $\omega_{1}$ and $\omega_{2}$ are to a very good approximation isosbestic points (this is illustrated in Fig.~\ref{Fig:isopoints} (Appendix~\ref{DLExtrapol}) by the enlarged view of one of the samples, which is indeed representative for all samples studied here). This implies that the slope of the intensity versus temperature is opposite in the region between the isosbestic points and the frequency regions below and above. While this applies to the general trend over the full 300 Kelvin span of temperatures, it does not apply to the sudden change of slope at $T_c$. In the examples that will follow we will see, that change of slope is either much smaller --~but with the same sign~-- below and above the inter-isosbestic region, or zero within the experimental accuracy.
The intra-isosbestic integral
\begin{eqnarray}
{E}_{C}^{iso}(T)&=&\frac{\hbar}{2\pi}\int_{\omega_1}^{\omega_2} L(\omega, T)
(1+2n_{\omega})
d\omega
\label{isocoulomb}
\end{eqnarray}
eliminates compensation of opposite temperature trends in different parts of the spectra, and provides for this reason the cleanest ({\em i.e.}~noise-free) representation of the temperature dependence of the experimental spectra in the region of the maximum of the loss function. 
This is also the spectral range where, faithful to the ``Willie Sutton principle''~\cite{Sutton1952} we anticipate the strongest saving of Coulomb energy. 
At the same time, since the definition of Coulomb energy requires integrating over all energy, it is also interesting to look in the other parts of the spectrum. 
To address these contributions to the Coulomb energy, the temperature dependence of the loss function integral below the lower isosbestic point and above the upper isosbestic points are reported for each sample in Appendix~\ref{fulldata}. 

To motivate the form in which we present our data we draw an analogy between $E_C(T)$ and the total internal energy $E(T)$.
In general, for a system in thermal equilibrium $E(T)$ is an increasing function of temperature, with a discontinuity in the slope at a second order transition. 
Usually one measure not the internal energy itself, but its temperature derivative $C_V=dE/dT$, {\em i.e.} the specific heat, which has a jump at a second order phase transition. 
Often in the cuprates the specific heat presents a $\Lambda$-like transition at $T_c$ rather than a jump. 
This type of broadening is often attributed to superconducting fluctuations, which must be properly accounted for when one tries to extract the condensation energy, $E_{cond}$, from the data~\cite{vandermarel2002}. 
Since the specific heat of a metal in the normal state is characterized by a linear temperature dependence ($E$ being proportional to $T^2$ as discussed in Appendix~\ref{Normaldata}), it is common practice to display the Sommerfeld coefficient $\gamma=T^{-1}C_V(T)$. 
In the present context, we concentrate our analysis on the corresponding quantity related to the Coulomb energy
\begin{equation}
\label{gamma_C}
\gamma_C^{iso}(T)=\frac{1}{T} \frac{d{E}_{C}^{iso}(T)}{dT}
\end{equation}
Since numerical evaluation of the temperature derivatives causes a strong amplification of the experimental noise, we can --~for this purpose~-- only use ${E}_{C}^{iso}(T)$, {\em i.e.} the loss function integrated between the two (sample-dependent) isosbestic points. 

\section{Loss function and partial Coulomb energy change through the superconducting phase transition}
\subsection{Qualitative features}\label{section4}
\begin{figure*}[t!!!]
\begin{center}
\includegraphics[width=1\textwidth]{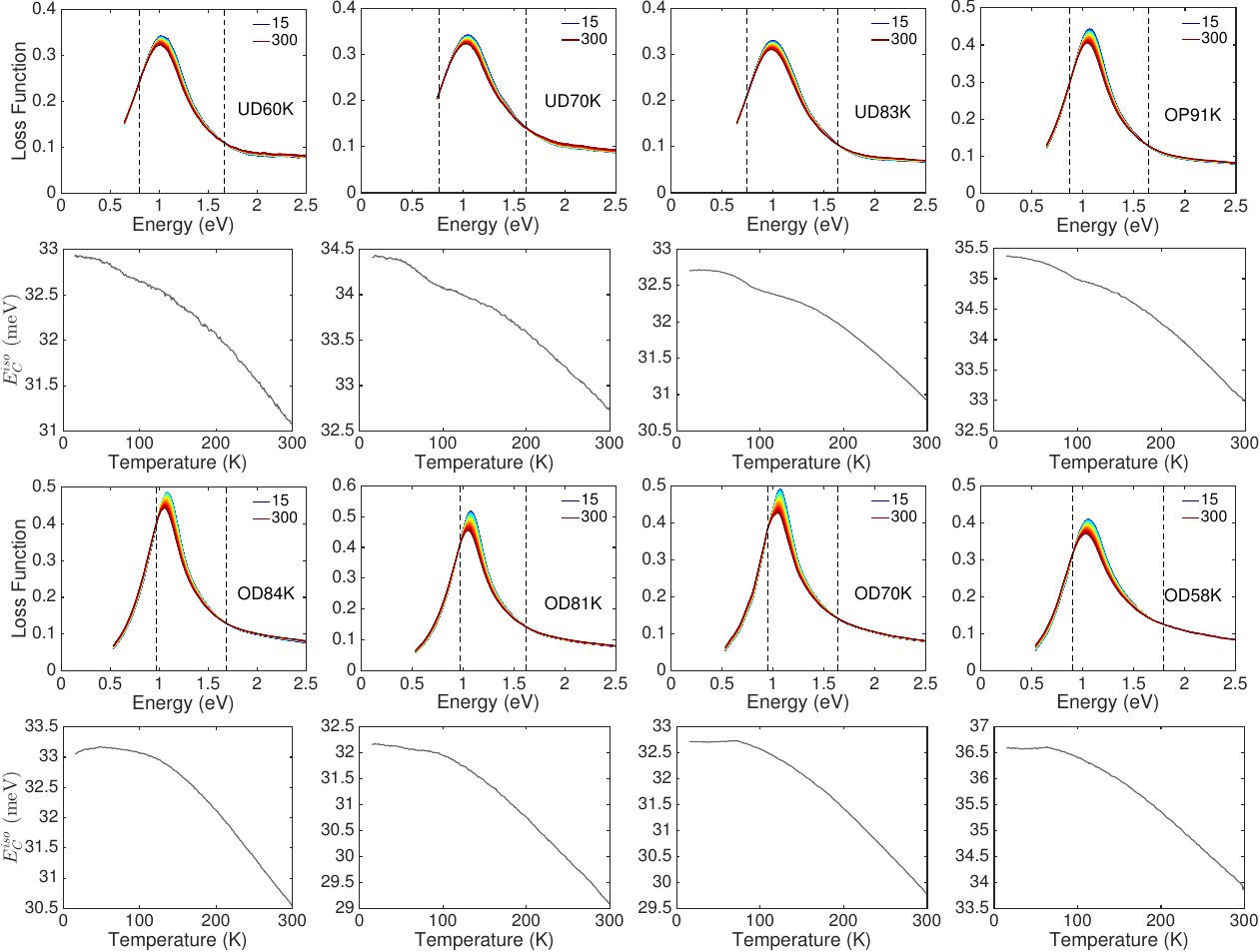}
\caption{\label{alldata} Loss function spectra for Bi2212 with carrier concentrations ranging from underdoped to overdoped for selected temperatures (lines 1 and 3). Temperature dependence of the integrated intensity, ${E}_{C}^{iso}(T)$, of the corresponding samples (lines 2 and 4).}
\includegraphics[width=1\textwidth]{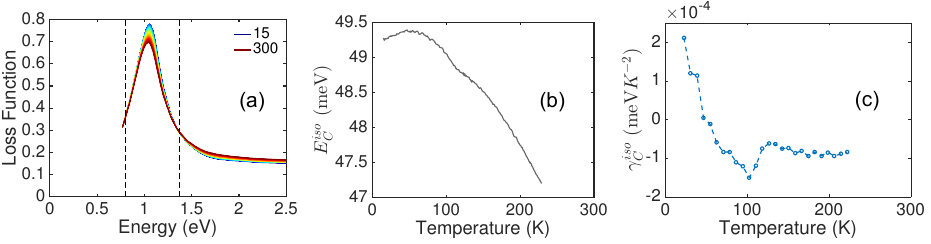}
\caption{\label{data2223}Optimally doped Bi2223 (a) Loss function spectra, (b) Temperature dependence of the integrated intensity ${E}_{C}^{iso}(T)$ and (c) Temperature dependence of $\gamma_C^{iso}(T)$.}
\end{center}
\end{figure*}
The loss function for the temperatures ranging from 15 to 300~K for all Bi2212 and Bi2223 samples is displayed in Fig.~\ref{alldata} and Fig.~\ref{data2223}.
The peak of the loss function corresponds to the plasma-resonance energy, which is at an energy slightly above 1~eV in all samples. 
Common to all samples of this study the intensity in the energy loss function increases gradually when cooling down, and gains approximately 5\% between 300~K and 15~K. 
There is a narrowing of the loss function peak, and a blueshift of about 5\%. 
For each of the samples all curves measured at different temperatures cross at two isosbestic points on either side of the maximum of the loss function. 

The real and imaginary parts of $-\epsilon(\omega)^{-1}$ for all samples are shown for different temperatures in Figs.~\ref{fig:Bi2212-60-UD}~--~\ref{fig:Bi2212-58-OD} of Appendix~\ref{fulldata}. 
In the same figures are also compared the temperature dependence of the intra-isosbestic loss function intensity ${E}_{C}^{iso}(T)$, and the loss function integrated from 0 to 2.5~eV (right panels of the second and third rows).
In the second (third) line of the first column are displayed the temperature dependences of the loss function integrals in the range below (above) the intra-isosbestic region.
For ease of comparison of the contributions from the different energy ranges these contributions are indicated on the same scale for a given sample. 

Based on these data we make the following global observations: In the first place the loss function integrals from 0 to $\hbar\omega_1$ (second row left panel), and $\hbar\omega_2$ to 2.5~eV (third row left panel) have very weak temperature dependence compared to the intra-isosbestic loss function integrals (from $\hbar\omega_1$ to $\hbar\omega_2$). In the second place the loss function integrals over the full (0 to 2.5~eV) range (third row, right panels) show by and large the same main features as the intra-isosbestic loss function integrals, while exhibiting stronger experimental noise. In view of these observations, and in the interest of the best possible signal-to-noise ratio, we concentrate for the details of the temperature dependence on the intra-isosbestic loss function integrals, ${E}_{C}^{iso}(T)$, which are displayed for all samples in the second and fourth lines of Fig.~\ref{alldata}. However, it is important to emphasize that this choice influences in no way our conclusions about the temperature dependence through $T_c$: The extended integrals from 0 to 2.5~eV, displayed in Figs.~\ref{fig:Bi2212-60-UD}~--~\ref{fig:Bi2212-58-OD} of Appendix~\ref{fulldata} show for all samples the same effects, both qualitatively and quantitatively, when $T$ is tuned through the superconducting phase transition.
The key aspects of the observed temperature dependencies, and the evolution thereof as a function of doping are the following: 

--~In the underdoped samples, we observe in ${E}_{C}^{iso}(T)$ an upward kink at $T_n$. An upward kink is also observed at $T_n$ for both optimally doped bilayer Bi2212 and for the tri-layer compound Bi2223. 

--~In the overdoped samples ${E}_{C}^{iso}(T)$ shows a downward kink below a temperature $T_p$. 

--~For Bi2223 ${E}_{C}^{iso}(T)$ turns downward below $60$~K. 
We speculate that this behavior has to do with the peculiarity, that the two outer planes and the inner plane of the trilayer compound have very different doping levels, as has been noticed with nuclear magnetic resonance~\cite{trokiner1991,kotegawa2001} and angle resolved photoemission (ARPES)~\cite{ideta2010}. 
Based on an analysis of the ARPES data it was estimated~\cite{ideta2010} that the outer planes are overdoped with $x=0.23$ holes per copper atom, while the inner plane is strongly underdoped with only $x=0.07$ holes per copper. 
This corresponds to an average doping $x=0.18$. 
We expect in this case a rich temperature dependence of ${E}_{C}^{iso}(T)$ of the coupled planes, combining aspects both of the underdoped and the overdoped side of the phase diagram, which indeed appears to be the case for the Bi2223 data (see Figs.~\ref{data2223} and~\ref{fig:Bi2223-110-OP}).
\begin{figure*}[tb]
\begin{center}
\includegraphics[width=1\textwidth]{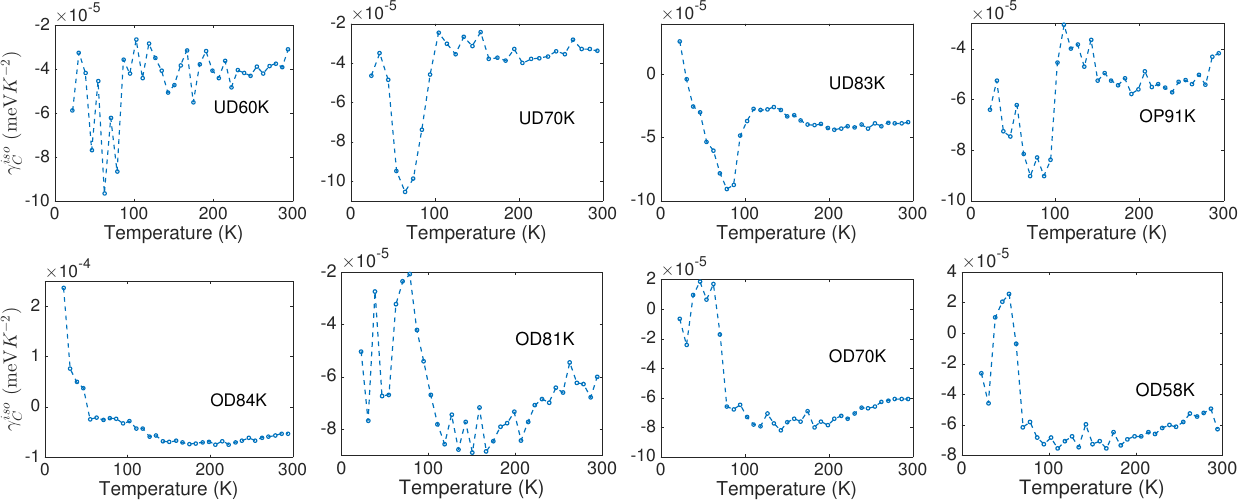}
\caption{\label{alldata2} Temperature dependence of $\gamma_C^{iso}(T)$ for Bi2212 with different carrier concentrations.}
\end{center}
\end{figure*}
Fig.~\ref{alldata2} displays $\gamma_C^{iso}(T)$ evaluated using Eq.~\eqref{gamma_C}. 
The main feature revealed by this quantity is the extremum close to $T_c$. 
Note the striking similarity of $\gamma_C^{iso}(T)$ in the overdoped samples to the Sommerfeld coefficient measured in a canonical superconductor. 
At the underdoped side the transition is broader, and $\gamma_C^{iso}(T)$ has a $\Lambda$-like appearance, similar to the Sommerfeld coefficient of underdoped cuprates obtained from specific heat experiments~\cite{Loram93,Loram94,vandermarel2002,Tallon01}. 
Of particular interest is the opposite sign of this extremum when comparing the underdoped and overdoped samples. 
This sign-change of the jump of $\gamma_C^{iso}(T)$ occurs for $p \approx 0.19$, which coincides with the point where a large body of experimental data indicates the closing of the pseudogap~\cite{Tallon01}. 
Recently Lee~\cite{lee2015} has obtained, from a formalism taking into account Umklapp processes, that the spectral weight in the loss function peak is enhanced when the materials switches from normal to superconducting. 
While this prediction agrees with the behavior that we observe for the underdoped samples, it is opposite to the effects seen on the overdoped side. 
Further theoretical studies should clarify the role and impact of the Umklapp processes as a function of doping.

In any case, the answer to question A (given as always the extrapolation assumption) is clear: the optical data are qualitatively consistent with the MIR scenario on the overdoped side of the phase diagram but not on the underdoped side.

\subsection{Quantitative considerations}
We now turn to question (B).
To quantify the change of Coulomb energy in the N--S phase transition we proceed in two steps: (i) We need to compare the measured data of the samples, which are superconducting, to the value without superconductivity. (ii) We need to estimate the average over the relevant sector of $\mathbf{q}$-space actually not of the loss function itself but of a related quantity (see below).
For point (i) we essentially need to extrapolate the normal state trend to zero. For the optimally doped sample with $T_c=91$~K the integrated loss function has a clean kink at $T_c$ (Fig.~\ref{alldata}, top right panel), and the behavior below $T_c$ behaves as $E_{C,s}^0(T)=E_{C,n}^0(T)+\Delta E_C^0[1-(T/T_c)^{\eta}]$, with $\Delta E_C^0=0.2$ meV and $1.5\lesssim \eta \lesssim 3$. The constant $\Delta E_C^0$ then represents the S--N difference of partial Coulomb energy for $q\sim 0$. For some of the samples the transition is less sharp, the normal state trend is less obvious, or a combination of these. To assure that the results of different dopings can be compared to each other, we calculate $\Delta E_C$ for each of the samples using the expression
\begin{equation}\label{def:C}
\Delta E_C^0=\eta^{-1}\left[\gamma_C(T_{2})-\gamma_C(T_{1})\right]T_c^2
\end{equation}
where $T_{1}\sim T_{c}$ and $T_{2}$ are the temperatures shown in Fig.~\ref{phasediag} characterizing the step in $\gamma_C(T)$. Since the choice $\eta=2$ provides for sample Op91 the expected result $\Delta E_C^0\sim 0.2$ meV, we use $\eta=2$ for all samples. The quantity $\Delta E_C^0$ forms a useful standard of comparison for the energies to be discussed below.

Point (ii) involves some rather delicate considerations concerning the meaning of the ``MIR scenario''.
To motivate them, let's note that the {\em total} Coulomb energy associated with wave vector $\mathbf{q}$ is given rigorously by Eq.~\eqref{E^C_q}. 
Thus, if we make our extrapolation assumption and assume for the moment that $L$ is not strongly dependent on the $c$-axis component of $\mathbf{q}$, the order of magnitude of the contribution to the total Coulomb energy from ``small $q$ and mid infrared $\omega$'' is simply given by $\Delta E_C^0$ multiplied by the fraction of the first Brillouin zone corresponding to the in-plane component of $\mathbf{q}$ being less than $q_0=0.31$~\AA$^{-1}$. 
This fraction is about $10 \%$, so that the resulting energy is about an order of magnitude smaller than the experimentally measured condensation energy.
However, this estimate is not in the spirit of the MIR scenario, which attributes the condensation energy to the saving of the Coulomb interaction energy {\em between the conduction electrons in the CuO$_2$ planes}; note that this interaction is screened by the core electrons, an effect which turns out to be quite significant quantitatively. 
We should therefore calculate this interaction energy (or rather the S--N difference in it) along the lines of Ref.~\onlinecite{Leggett1999a,Leggett1999b} or via a related ``3D'' approach; see appendix \ref{C-axis_contribution}. 
To object that while the fraction of $\Delta E_C^0$ thus obtained may be quite large, the total Coulomb energy saving associated with small $q$ and mid infrared $\omega$ is much smaller, is no more compelling than would be an objection to the physical relevance of the ``kinetic energy sum rule'' ({\em cf.} section \ref{section2}) on the grounds that it does not take account of the change of kinetic energy of the core electrons. 
In both cases we are exploring the situation {\em at the level of a model}, and the outcome may look qualitatively different from the exact Dirac-level picture.

The quantity whose difference in the N and S phases we want to estimate is given by Eq. (5.1.1) of Ref.~\onlinecite{Leggett1999b}, with the factor $K(\omega)$ given by Eq. (4.1.4) of that reference; for the special case, relevant to Bi-2212, of $n=2$ the resulting expression for the ``mid infrared''  (MIR) scenario is 
\begin{align}\label{eq14}
&E_{C}^{mir}=\frac{a^2}{8\pi^2}\int_0^{q_0}qdq
\int_0^{\infty} d\omega \nonumber \\
&\sum_{p=\pm 1}
\mbox{Im}
\frac{-1}{1+
\left[1+pe^{-qd}\right] (q\bar{d}/2)\left[\epsilon(\omega)-\epsilon_b\right] /\epsilon_{sc}
}
\end{align}
In the expression~\eqref{eq14} the quantity $\bar{d}$ is the {\em mean} plane spacing (7.8 $\AA$ for Bi-2212), $\epsilon_{sc}$ is the factor, assumed frequency-independent, by which the Coulomb interaction between the conduction electrons in the CuO$_2$ planes is screened by the ionic cores, and $\epsilon_b$ is the ``background'' dielectric constant, arising from not only the in-plane Cu and O ions and the intercalated Ca but also from the ions in the ``charge-reservoir'' layers, which has to be subtracted from the experimentally measured $\epsilon(\omega)$ to get the conduction-electron contribution. 
In evaluating the expression~\eqref{eq14} we have taken $\epsilon_{sc}=\epsilon_{b}=4.5$ (see appendix \ref{Bound}). The formula~\eqref{eq14} is based on a ``2D'' treatment (``method 1'') in which one regards the planes as separate from the 3D ``background'' ignores the effect of inter-planar Coulomb interactions, which are important only for $qd_c \lesssim 1$ , {\em i.e.} $q\lesssim 0.065$ \AA$^{-1}$. 
{
In appendix~\ref{C-axis_contribution} we generalize the expressions to finite inter-planar Coulomb interaction. For the present parameters the different methods, {\em i.e.} modeling as a 3D stack of $\delta$-layers (Eq. \eqref{c-axis_corr}), or as a bilayer (Eq. \eqref{eq14}) result in the same estimate of the momentum integrated Coulomb energy (see Fig.~\ref{fig:caxis}), showing that the finite $\mathbf{q}$ extrapolation effectively corresponds to multiplying the $q=0$ value with a factor $F=0.42$. Using the same method we obtained $F$ for all other dopings, giving $F\sim 0.4$ with weak sample-to-sample variations. Together with the output of Eq. \eqref{def:C} this yields the MIR-regime Coulomb energy $\Delta E_{C}^{mir}=F\Delta E_C^0$.}

\begin{figure}[t!!!!]
\centerline{
\includegraphics[width=\columnwidth,clip=true]{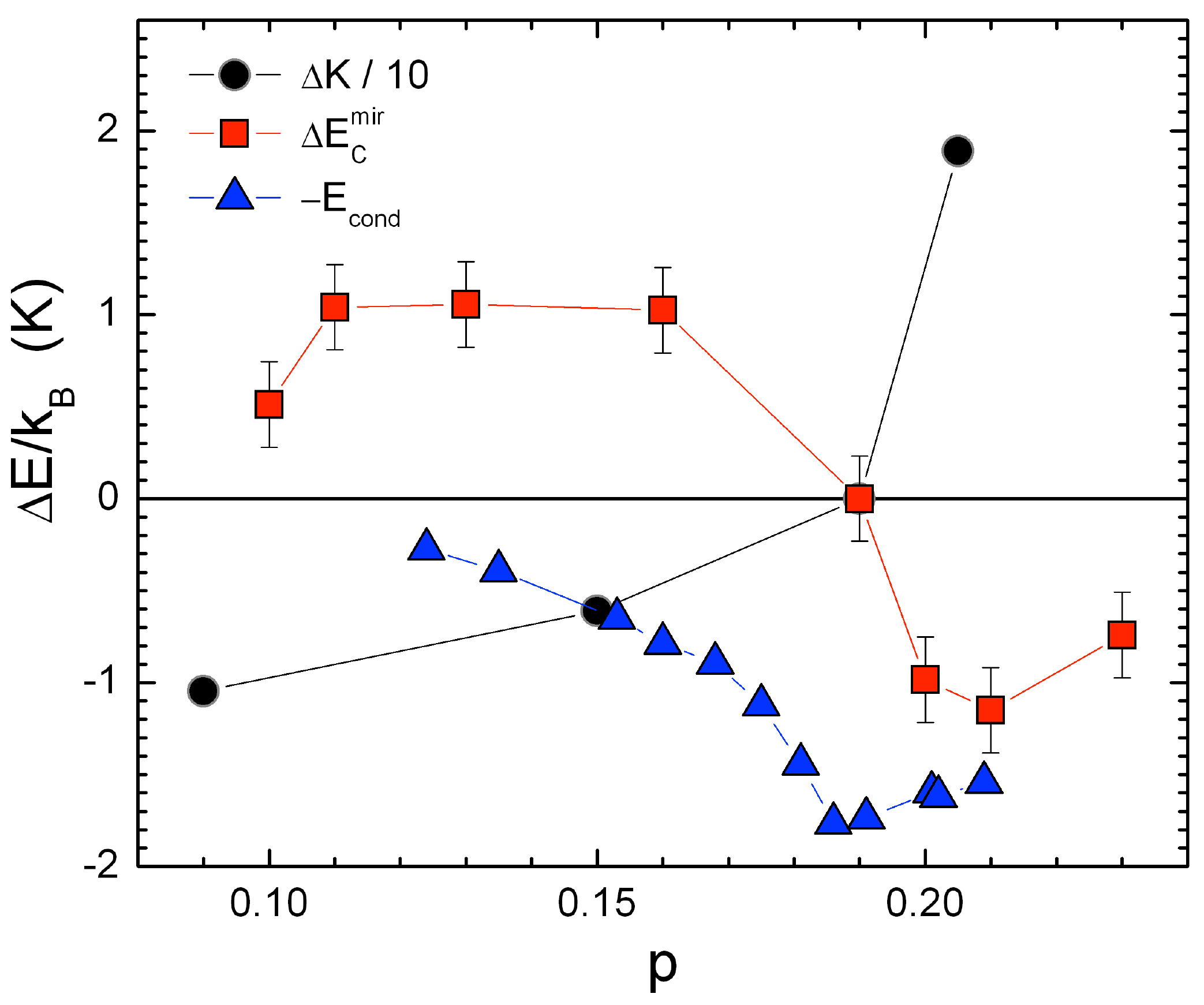}
}
\caption{\protect 
S--N difference of the $\mathbf{q}$-integrated Coulomb energy, $\Delta{E}_{C}^{mir}$ , together with the total energy difference $-E_{cond}$ (data reproduced from Ref.~\onlinecite{Loram2000}, with original units converted to the present ones for the sake of comparison), and band-energy difference, $\Delta{K}$~\cite{carbone2006b}. }
\label{fig:energies}
\end{figure}

The doping dependence of $\Delta E_{C}^{mir}$ is presented in Fig.~\ref{fig:energies}. This figure constitutes the central result of this study. 
In addition are shown the S--N difference of the total energy $-E_{cond}$ ({\em i.e.} minus the condensation energy ) obtained from specific heat~\cite{Loram2000}, and $\Delta{K}$ from the sum rule Eq.~\eqref{K-sumrule}, representing the difference in band-energy (``kinetic energy'') between superconducting and normal state~\cite{carbone2006b}. 
Important for the interpretation of the data is the comparison of the absolute values of the energies involved: The Coulomb energy change is in the range $-1$ to $1$~K, the condensation energies are in the range $0$ to $2$~K, and the kinetic energy changes are in the range $-10$ to $20$~K. 
We note (a)~that the general trend of the MIR-regime Coulomb energy as a function of doping, in strong contrast to that of the ``kinetic'' energy, is similar to that of the total condensation energy (b)~that in the overdoped region ($p>0.19)$ it can contribute to the latter, though it obviously cannot be the whole cause of superconductivity.
This answers the question stated in the introduction of this paper, namely whether the saving of the inter-conduction electron energy, and specifically the part associated with long wavelengths and mid-infrared frequencies drives the superconducting transition in the cuprates. 
The answer to this question is ``it is an important factor in the energy balance, but not the only factor driving the mechanism of pair formation''. 
However, to obtain this answer we have assumed, rather than tested experimentally, the momentum dependence predicted by the MIR model. 
Future experimental studies using electron energy loss spectroscopy are needed to test the prediction about the momentum dependence of the partial Coulomb energy. 

These results do not exclude the possibility that the Coulomb energy in the $(\pi,\pi)$ region is important for stabilizing the superconducting state. On the microscopic level this can involve super-exchange, interaction mediated by the virtual exchange of spin-fluctuations, or other many-body effects involving the Coulomb potential.
The results in Fig.~\ref{fig:energies} show a striking similarity with the theoretical results of Gull and Millis, shown in the lower panel Fig.~2 of Ref.~\onlinecite{gull2012}. These results were obtained using the dynamical cluster approximation (DCA) version of dynamical mean-field theory~\cite{maier2005} for the Hubbard model, taking for the on-site repulsive interaction the value $U=6t$. In G\&M the kinetic energy and interaction energy refer to the expectation values of the corresponding two terms (the only ones) of their Hamiltonian. Particularly striking is the agreement with the change of sign of $\Delta {K}$ and $\Delta E_C$, which follows the same trends both in experimental and computational data. The doping level where the sign-change occurs is different ($p=0.08$ in Fig.~2 of Ref.~\onlinecite{gull2012}), which is not surprising considering that the band structure details are different between the theoretical model and in Bi2212. The most important difference with the numbers shown in Fig.~\ref{fig:energies} is, that the interaction energy in the latter refers to the long-wavelength limit of the Coulomb energy, which in reality diverges as $e^2/q^2$ whereas the Hubbard interaction is independent of $q$. It would be interesting to analyze G\&M's method to see how the saving of interaction energy depends on $q$, and to study extended versions of this model to include the $e^2/q^2$ dependence in the long-wavelength limit. G\&M find in their numerical data no indication that the nonsuperconducting pseudogap state has any significant pairing correlations. The sign-change seen both in the theoretical and in the experimental data is however compatible with a competition between the pseudogap, present at the underdoped side, and superconductivity. 

\section{Higher-frequency and higher-temperature effects}
Due to the fact that the Coulomb energy can not be treated perturbatively, the question as to which energy regime contributes most to the condensation energy is very difficult to answer theoretically. 
The emphasis on the photon energy range from $0.6-1.8$~eV was made primarily because this is the most obvious regime where the normal state loss function is both substantial and reasonably system-independent. 
This assumption may have been too restrictive since even if inter-band transitions may dominate the optical response above 2~eV, these bands are probably mostly associated with the CuO$_2$ planes and/or the apical oxygens which are common to the various cuprates. For example, Eqs.~\eqref{eq14} and~\eqref{c-axis_corr} implicitly attribute a fraction $1-1/\epsilon_{sc}$ of $\Delta E_{C}^{mir}$ reported in Fig.~\ref{fig:energies} to the range of inter-band transitions of the loss function (for details see Appendix~\ref{Bound}).
These considerations lead to hypothesis C: An important contribution to the Coulomb energy saving originates in the energy range of charge-transfer transitions above the free carrier plasma-frequency.

We note one further point, namely that originally it was assumed~\cite{Leggett1999a} that the saving of the Coulomb energy would sets in at $T_c$, however, more recently, and in the light of the earlier rounds of this experiment, two of us~\cite{AJL&DP2} have considered the possibility, in a more general scenario, of the onset of a drop in the Coulomb energy at some temperature fairly well above $T_c$ -- crudely speaking because of preformed Cooper pairs~\cite{Emery1995} -- in a temperature range where other experiments seem consistent with the onset of local Cooper pairing~\cite{wang2005,wang2006,gomes2007,tallon2011,dubroka2011,kondo2011,uykur2014,kondo2015}.
These considerations lead to hypothesis D: The experimental results obtained up to this point could be compatible with a gradual process whereby upon lowering the temperature finite-range pair-correlations become progressively facilitated by a saving of Coulomb energy. 

\subsection{Hypothesis C: Susceptibility well above the plasma frequency.}

In order to investigate the temperature dependence of the loss-function well above the main plasmon peak, we analyze the Coulomb energy integrated from zero up to 2.5~eV (in view of the large amount of noise in some of the samples in the range $2.5-3.1$~eV, we omit this range from the integral). 
For frequencies below the frequency range of the instrument (0.6~eV) the loss function was obtained from a Drude-Lorentz fit to the experimental data of both real and imaginary parts of $\epsilon(\omega)$ {\em simultaneously}. 
While this does not provide fine details such as optical phonons, the real {\em and} imaginary parts of $\epsilon(\omega)$ in the range 0.6 to 3.1~eV narrowly constrain the range of possible values of the loss function integral between 0 and 0.6~eV, as revealed by the low noise level of the temperature dependence shown in the middle left panels of Figs.~\ref{fig:Bi2212-60-UD} --~\ref{fig:Bi2212-58-OD}. In Appendix~\ref{DLExtrapol} we demonstrate the validity of this procedure for the example of Bi2223 by comparing the extrapolation with experimental data in the low frequency range. 
Comparing the middle left panels with the middle right panels of all samples, we see that this energy range has a smooth and weak temperature dependence; in particular the trends at the superconducting transition are not affected if we add this part to the integral. 
The energy range beween the upper isosbestic point and 2.5~eV is more affected by instrument noise. 
The general trend as a function of doping and temperature is, that the temperature dependence approaches increasingly a $T$-linear behavior towards overdoping, and this trend is more pronounced when we extend the integration range from upper isosbestic point to 2.5~eV.

Taking the sum of all three zones results in curves for optimally doped and weakly underdoped samples showing maxima at around 150~K. These maxima are a direct consequence of adding the intra-isosbetic $B-AT^\alpha$ with $\alpha\sim 2$ and the contributions from zones 1 and 3 of the form $C+DT^\beta$, where $\beta\sim 1$. 
In view of our current lack of understanding of these temperature-dependences, it is unclear whether the maxima have any significance within the MIR scenario.

All in all the verdict regarding hypothesis C is: The energy range of charge-transfer transitions above the free carrier plasma-frequency contributes significantly to the temperature dependence of the normal state. 
Insofar the changes across the superconducting phase transition are concerned, the contribution to the Coulomb energy in the narrow range of about 1eV above the plasma frequency is too small to be observable in most of our samples. 
\subsection{Hypothesis D: Search for fluctuations far above~$T_c$.}
In relation to hypothesis D we point out recent experiments~\cite{wang2006,cyrchoiniere2015} indicating that in the cuprates superconductivity competes with various different states of matter. In particular a fluctuating charge-density wave (CDW) has been observed in the cleanest high-$T_c$ system YBa$_2$Cu$_3$O$_{6+y}$ (YBCO) for $T^\star > T_{CDW} > T_c$~\cite{sidis2013,wu2011,Wu13,hashimoto2010,Blanco14,Hucker14}, but also in La$_{2-x}$Sr$_x$CuO$_4$~\cite{Croft14}, which in principle may give signatures in the temperature dependence of the Coulomb energy similar to those observed for a superconducting phase transition. 
Even if Bi2212 and YBCO have structural differences, an analogy may exist, in particular because charge order is also observed in Bi2212 by STM measurements~\cite{Lawler10,Parker10}. 

Such phenomena might very well also occur in the Bi2212 and Bi2223 systems which are the subject of the present experimental study. However, it is in principle difficult to tell apart fluctuating charge density waves and fluctuating superconducting order on the basis of the temperature trends observed in the optical data. Consequently, the weak temperature features that we will discuss in the remainder of this chapter may be attributed to either of these two, as well as other forms of fluctuating order.
\begin{figure}[htbp]
\begin{center}
\includegraphics[width=\columnwidth]{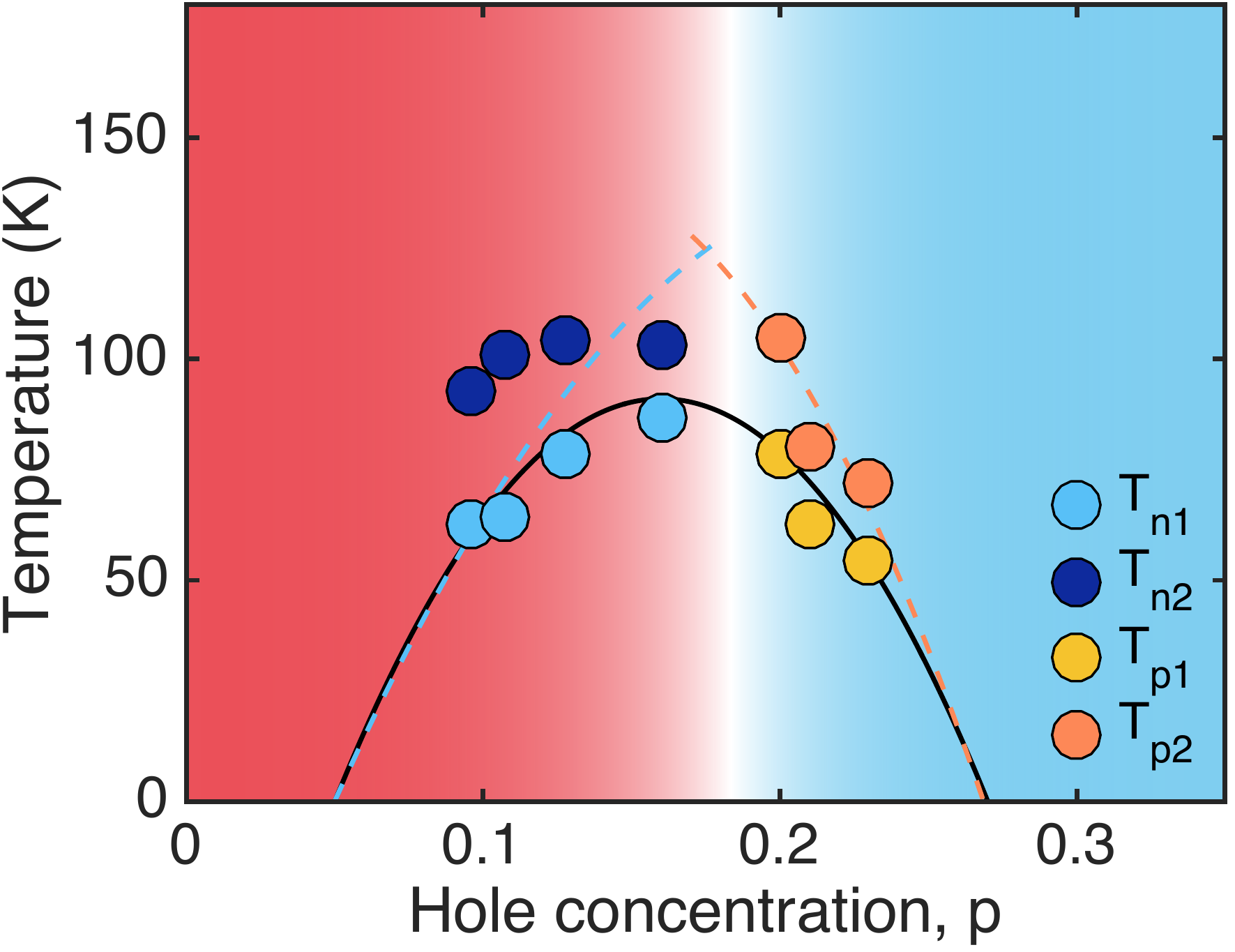}
\caption{Phase diagram summarizing the temperature characterizing the step in in $\gamma_C^{iso}(T)$: The temperature of the extremum, coinciding with the superconducting phase transition ($T_{n1}$/$T_{p1}$), and the upper limit of the step ($T_{n2}$/$T_{p2}$).}
\label{phasediag}
\end{center}
\end{figure}
Fig.~\ref{phasediag} summarizes all values of $T_{n1}$, $T_{n2}$, $T_{p1}$ and $T_{p2}$ as a function of hole concentration for all Bi2212 samples and highlights the change of nature of the $T_{n1}$ extremum in $\gamma_C^{iso}(T)$. 
It is clear that $T_{n1}$ and $T_{p1}$ (light blue and gold circles) can be associated to the critical temperature $T_c$ (defined by the empirical Tallon-Presland relation). 
The transition seen in $\gamma_C^{iso}(T)$ is broadened, especially on the underdoped side, and the resulting curves have a $\Lambda$-like appearance, similar to the Sommerfeld coefficient of underdoped cuprates obtained from specific heat experiments~\cite{Loram93,Loram94,vandermarel2002,Tallon01}. This aspect is most likely the consequence of fluctuations of the superconducting order above the critical temperature. 
Concerning $T_{n2}$ (navy blue circles), the dome shape that it defines, peaked around $p = 0.12$, strongly suggests an analogy between this energy scale and the one determined by Nernst effect measurements and associated with a range of temperatures where pair-correlations persist above $T_c$, but where the phase of the order parameter is strongly fluctuating~\cite{wang2006}.
Additional indications that in Bi2212 pair-correlations persist far above $T_c$ come from diamagnetism~\cite{wang2005}, scanning tunneling spectroscopy~\cite{gomes2007}, specific heat~\cite{tallon2011} and angle resolved photo-emission~\cite{kondo2011,kondo2015} experiments. 
The sign-change of the jump of $\gamma_C^{iso}(T)$ when going from underdoped to overdoped samples occurs for $p \approx 0.19$. 
This doping level coincides with the point where a large body of experimental data indicates the vanishing of the pseudogap~\cite{Tallon01}. 

\begin{figure}[htbp]
\begin{center}
\includegraphics[width=\columnwidth]{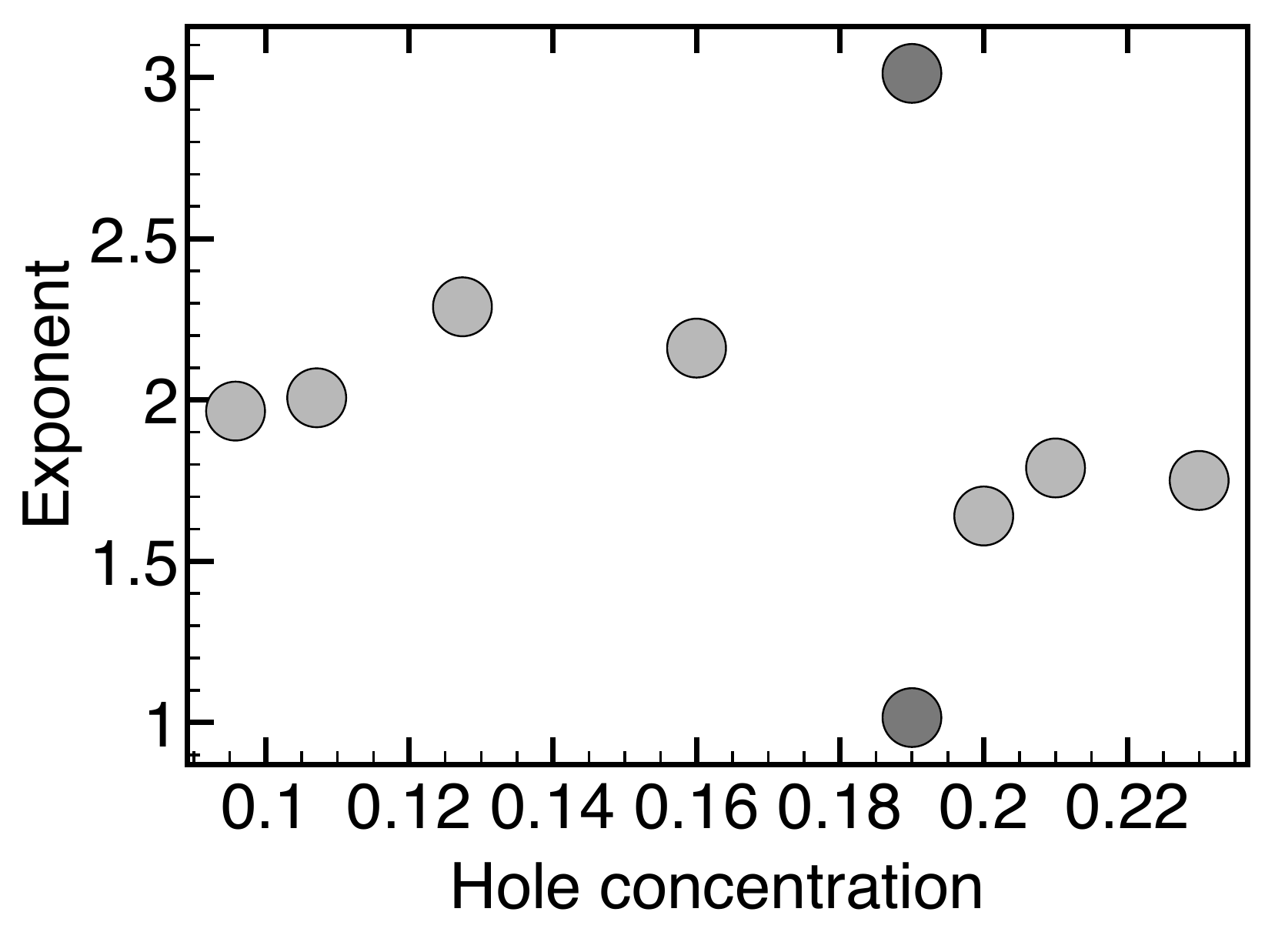}
\caption{Exponents of the equation ${E}_{C}^{iso}(T)=B+AT^\alpha$ obtained from fitting the normal state temperature dependence of the intra-isosbestic integrated loss-function intensity, shown in Figs.~\ref{fig:Bi2212-60-UD}~--~\ref{fig:Bi2212-58-OD}. 
The two values indicated for the 19 percent doped sample (dark gray) refer to different temperature regions.}
\label{fig:exponents}
\end{center}
\end{figure}

We now turn our attention to the temperature dependence farther above $T_c$. The red curves in Figs.~\ref{fig:Bi2212-60-UD}~--~\ref{fig:Bi2212-58-OD} are phenomenological fits of the function ${E}_{C}^{iso}(T)=B+AT^\alpha$ to the data above $T_c$ (where $B$ and $A$ are some constants). 
Such a power law temperature dependence, with $\alpha\sim 2$, is by and large described by a constant value of $\gamma_C^{iso}(T)$. 
This behavior finds a natural explanation in the temperature dependence of the free-carrier response at high frequencies, as explained in Ref.~\onlinecite{norman2007} (see Appendix~\ref{Normaldata}). 
On a qualitative level even from visual inspection of these temperature dependences for different dopings we can already conclude that for the overdoped samples the normal state evolution is less curved than for the underdoped samples. 
To substantiate this qualitative observation, the evolution from underdoped to overdoped is best illustrated by a plot of the fitted exponents, $\alpha$, as a function of the hole carrier concentration, shown in Fig.~\ref{fig:exponents}. 
The temperature dependence between $T_c$ and 300~K changes progressively as a function of doping. For 19 percent doping the value of the exponent depends strongly on the temperature range fitted, which should be taken as an indication that the temperature dependence is not algebraic for this doping. Interestingly this is also where the kink at $T_c$ changes from positive to negative.
The general trend is, that the curvature is stronger for underdoped samples, but the exponent returns to the value 2 for the lowest doped material. 
It should be emphasized, that for all dopings the evolution as a function of temperature above $T_c$ is gradual.
Nonetheless, comparing the (not too) underdoped and optimally doped samples, the observed behavior may be an indication that upon lowering temperature already far above $T_c$ the Coulomb energy of the underdoped samples flows to a lower value than for the overdoped ones. 
One can then speculate that this is the result of fluctuations of some kind of order parameter (Charge Density Wave, pairing, or other) far above $T_c$ in the underdoped cuprates.

On a more detailed level, for samples UD70K, UD83K, OP91K and possibly Bi2223 we observe in $\gamma_C^{iso}(T)$ (which is negative) a gradual drop for $T>T_{n1}$ which approaches the constant value for temperatures around 150~K
~\footnote{The slight undershoot of the data as compared to the fitted power law appears to vanish gradually as a function of increasing temperature. 
Consequently the fitted curve and the experimental data appear to merge at a temperature determined by the instrument noise, which obviously has no particular significance with respect to the state of matter of the material.}.
This hints at an accelerated saving of Coulomb energy when cooling down below $\sim 150$~K compared to the trend at higher temperatures.

All in all the verdict regarding hypothesis D is: We observe possible indications of rather gradual changes of the Coulomb energy in the normal phase, which are possibly associated to the presence of a fluctuating order of some kind in a region above $T_c$. First of all, the steps in $\gamma_C(T)$ are broadened, especially on the underdoped side, which is almost certainly indicative of a region of fluctuating superconducting order. Secondly, more subtle and gradual bending of the temperature dependence, especially on the underdoped side, may be due to fluctuations of unknown origin, which disappear gradually as a function of temperature.
\section{Summary and conclusion}
To summarize, we have measured the evolution as a function of temperature and doping of the loss function spectra in the infrared-visible spectral range of double- and triple layer bismuth cuprates. 
Our experiments indicate that for the overdoped samples the superconducting phase transition is accompanied by a saving of the Coulomb interaction energy, on the underdoped side there is an increase of the Coulomb energy below $T_c$, and the change of Coulomb energy for $q< 0.31$~\AA$^{-1}$ is about the same size as the condensation energy. { \color{black} This state of affairs calls for studies with other experimental techniques, in particular electron energy loss spectroscopy, to explore the momentum dependent structure of these phenomena.} Departure of a $T^2$ dependence of the measured loss-function data indicates a corresponding temperature dependence of the density-density correlations. Unambiguous assignment to a precursor of superconducting pairing, to another type of correlation, or neither of these two, is not possible at this stage.
The S--N difference of the Coulomb energy has similar doping dependence as the total condensation energy. While the latter is in the range of 0 to 2 K  per CuO$_2$ unit, the Coulomb energy varies between -1 and 1 K. Consequently, while it cannot be the whole cause of superconductivity, the Coulomb energy is a major factor in the total energy balance stabilizing the superconducting state. The experiments presented here demonstrate that it is in principle possible to determine the subtle changes of Coulomb correlation energy associated with a superconducting phase transition, and constitute a promising first step in the experimental exploration of the Coulomb correlation energy as a function of momentum and energy.
\begin{acknowledgments}
The authors thank Mehdi Brandt, J\'er\'emie Teyssier and Spiros Zanos for technical support. This project was supported by the Swiss National Science Foundation (project 200021 -- 162628). 
The work at Brookhaven is funded through DOE Contract No. DE-SC00112704. 
The work at the University of Illinois at Urbana-Champaign was supported as part of the Center for Emergent Superconductivity, an Energy Frontier Research Center funded by the U.S. Department of Energy, Office of Science, Office of Basic Energy Sciences under Award Number DE-AC02-07CH11358 
\end{acknowledgments}
\appendix
\begin{figure}[t!!!]
\begin{center}
\includegraphics[width=1.00\columnwidth]{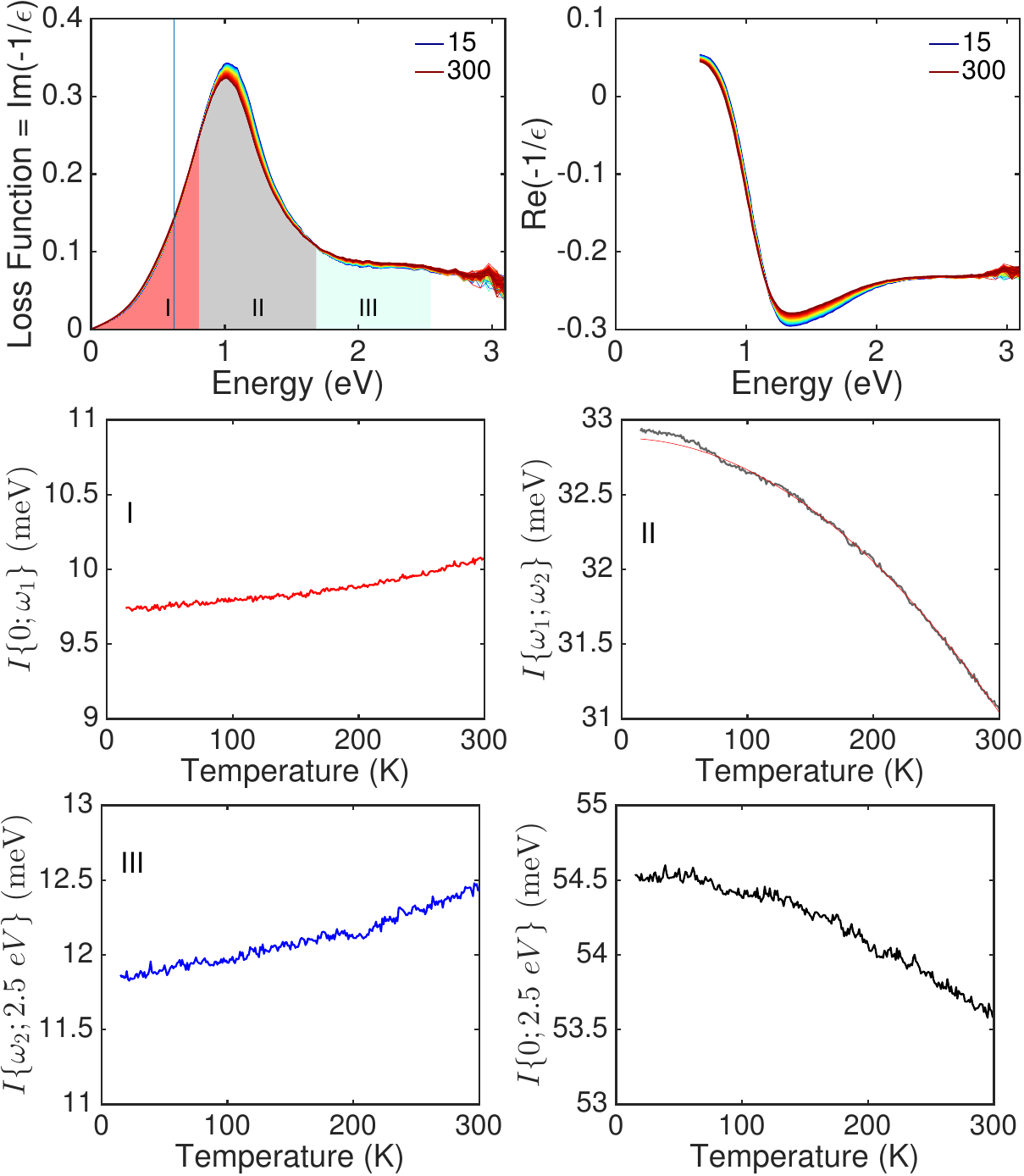}
\caption{Loss function and loss function integrals of sample Bi2212-60-UD. 
The value of the fitted exponent $\alpha$ is $2.0$ (right middle panel). 
Panel details are further specified in the main text.}
\label{fig:Bi2212-60-UD}
\end{center}
%
%
\begin{center}
\includegraphics[width=1.00\columnwidth]{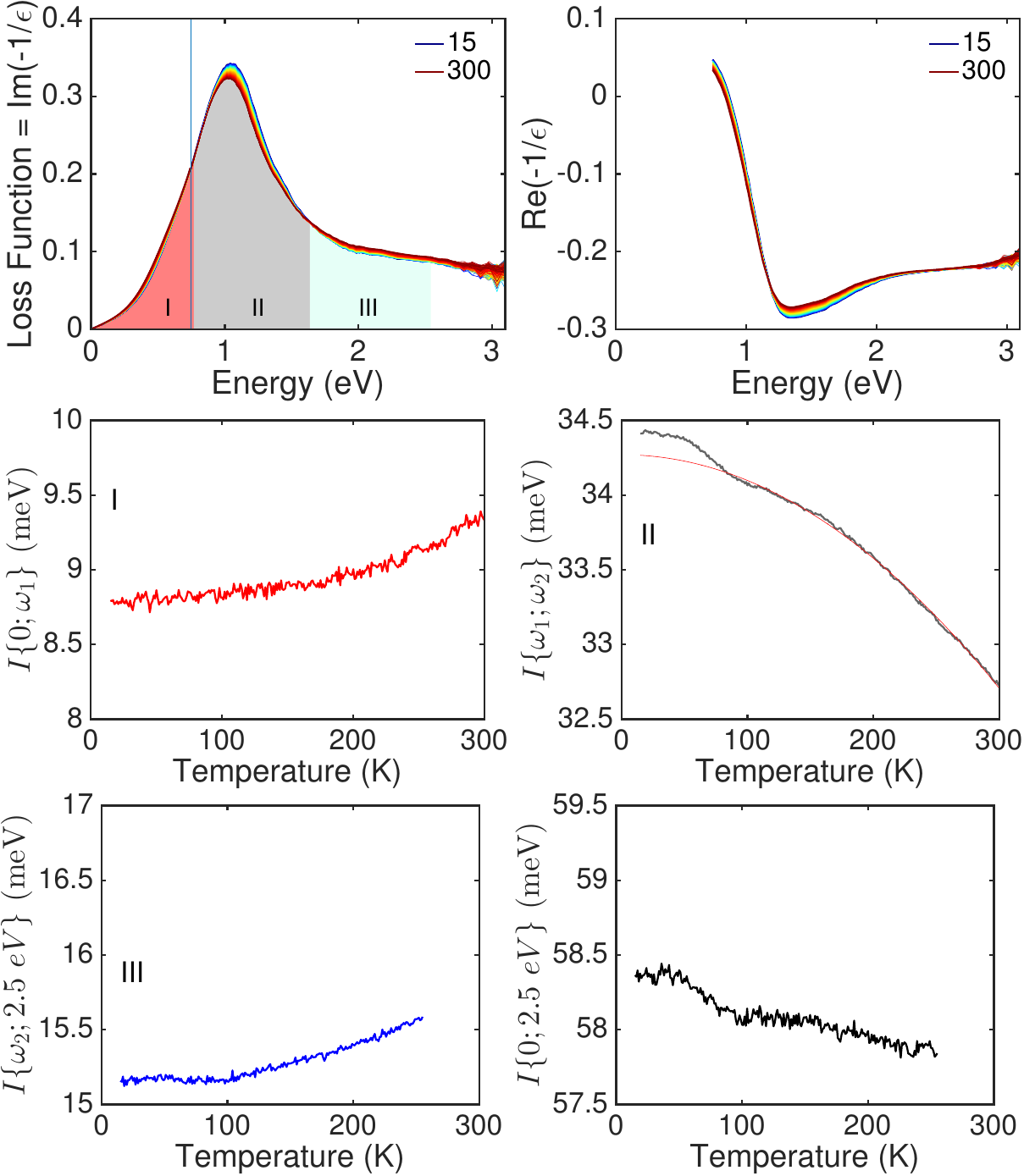}
\caption{Sample Bi2212-70-UD. Caption details as in Fig.~\ref{fig:Bi2212-60-UD}. 
Because the signal for frequencies in zone 3 was not sufficiently stable above $256$~K, the temperature range of the panels on the third line is limited below $256$~K. 
The value of the fitted exponent $\alpha$ is $2.0$. }\label{fig:Bi2212-70-UD}
\end{center}
\end{figure}
\begin{figure}[t!!!]
\begin{center}
\includegraphics[width=1\columnwidth]{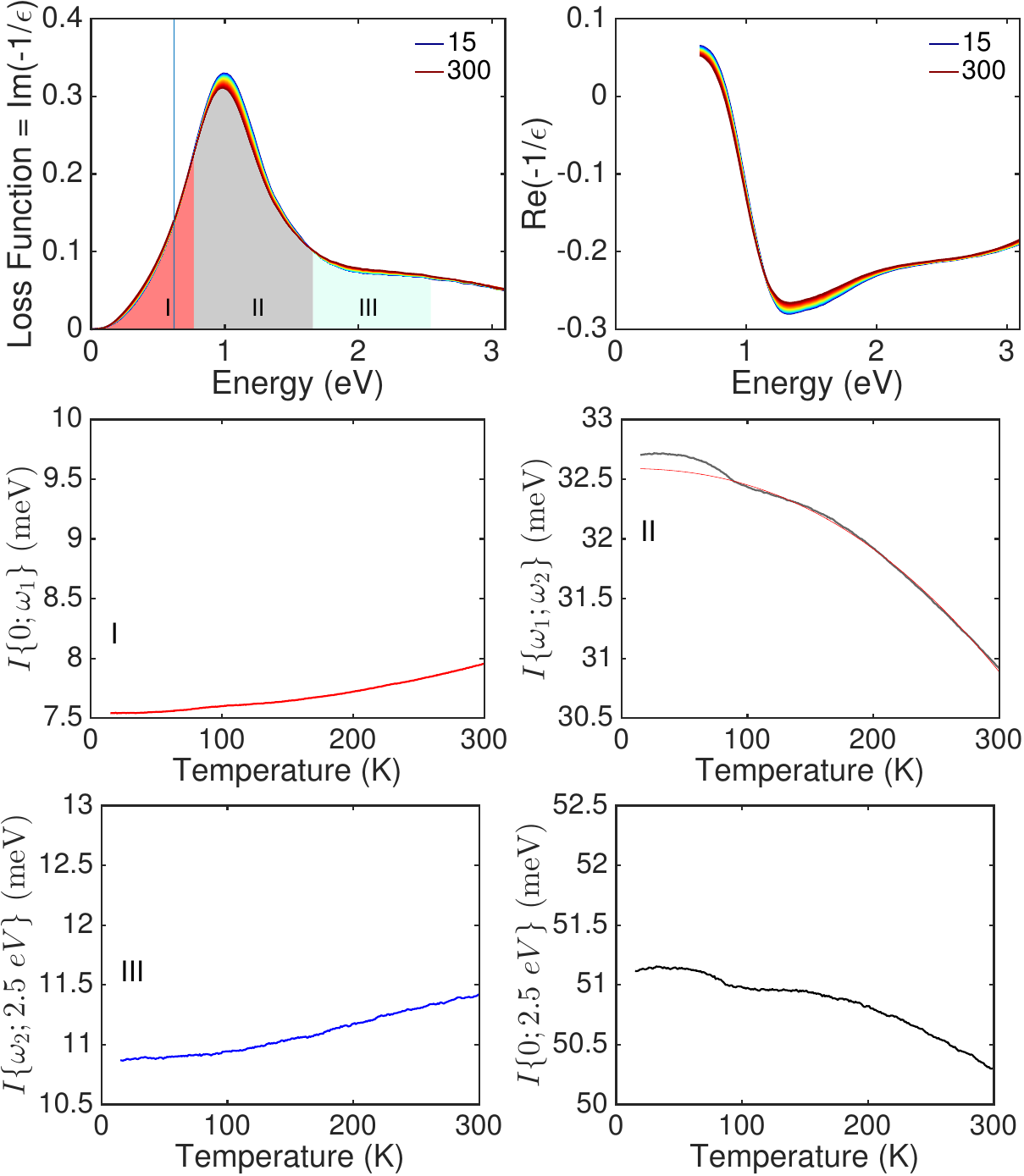}
\caption{Sample Bi2212-83-UD. 
Caption details as in Fig.~\ref{fig:Bi2212-60-UD}. 
The value of the fitted exponent $\alpha$ is $2.3$. }\label{fig:Bi2212-83-UD}
\end{center}
%
%
\begin{center}
\includegraphics[width=1.00\columnwidth]{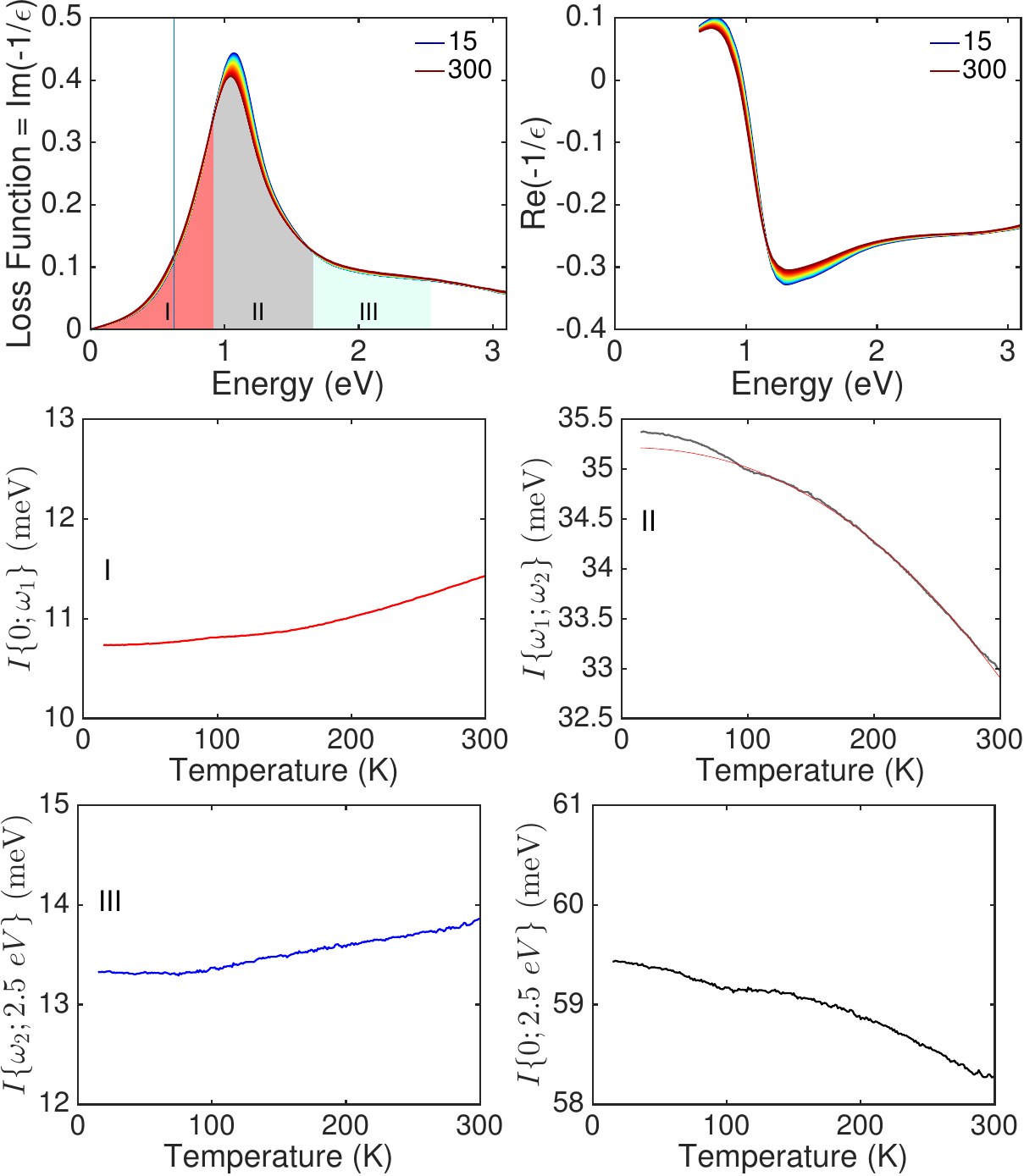}
\caption{Sample Bi2212-91-OpD. 
Caption details as in Fig.~\ref{fig:Bi2212-60-UD}. 
The value of the fitted exponent $\alpha$ is $2.2$. }\label{fig:Bi2212-91-OP}
\end{center}
\end{figure}
\begin{figure}[b!!!]
\begin{center}
\includegraphics[width=1.00\columnwidth]{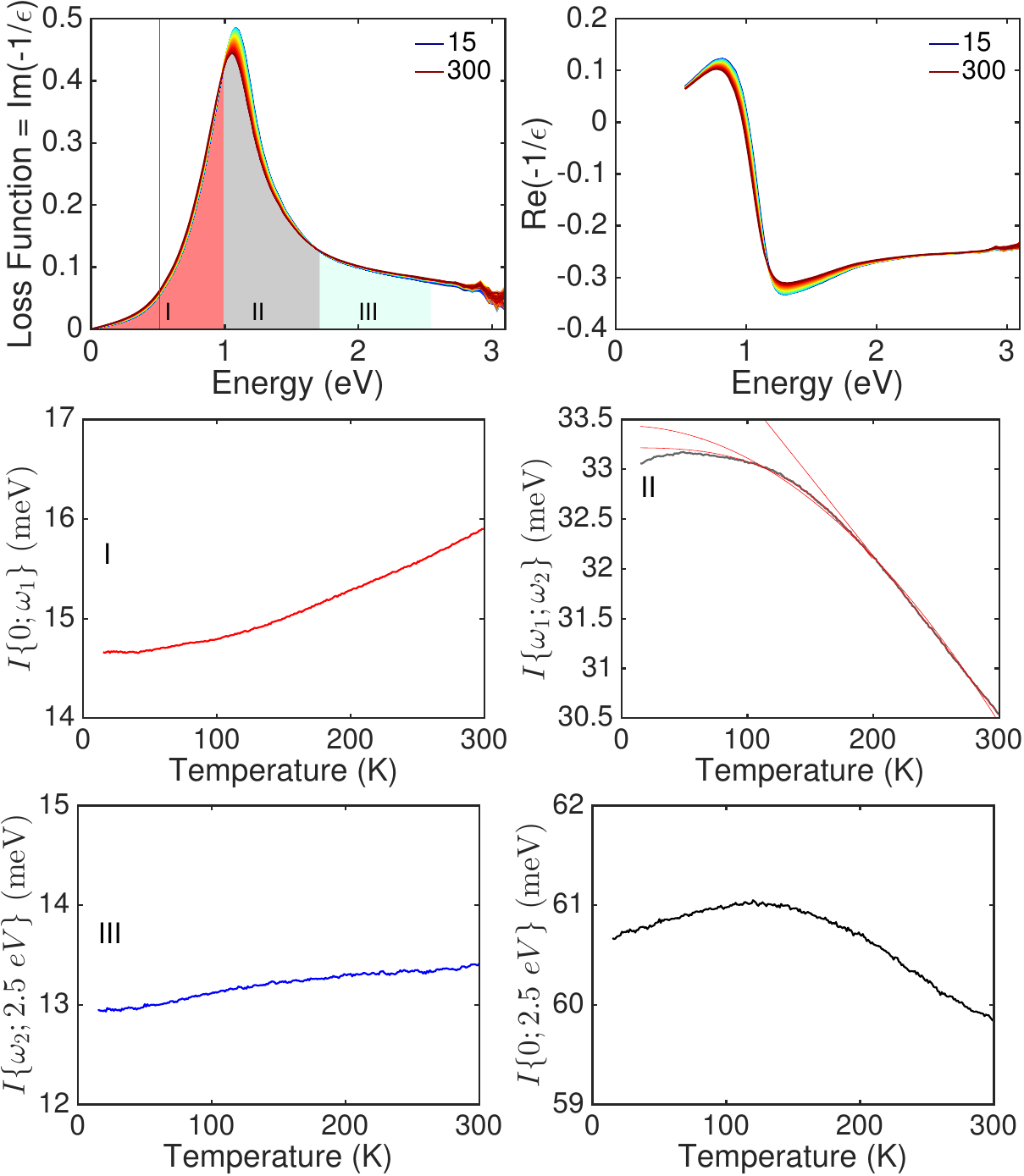}
\caption{Sample Bi2212-84-OD. 
Caption details as in Fig.~\ref{fig:Bi2212-60-UD}. 
The value of the fitted exponent $\alpha$ is $2$ when fitting down to $T_c$ and $1$ and $3$ when splitting into $T_c-200$~K and $200-300$~K windows.}\label{fig:Bi2212-84-OD}
\end{center}
%
\begin{center}
\includegraphics[width=1.00\columnwidth]{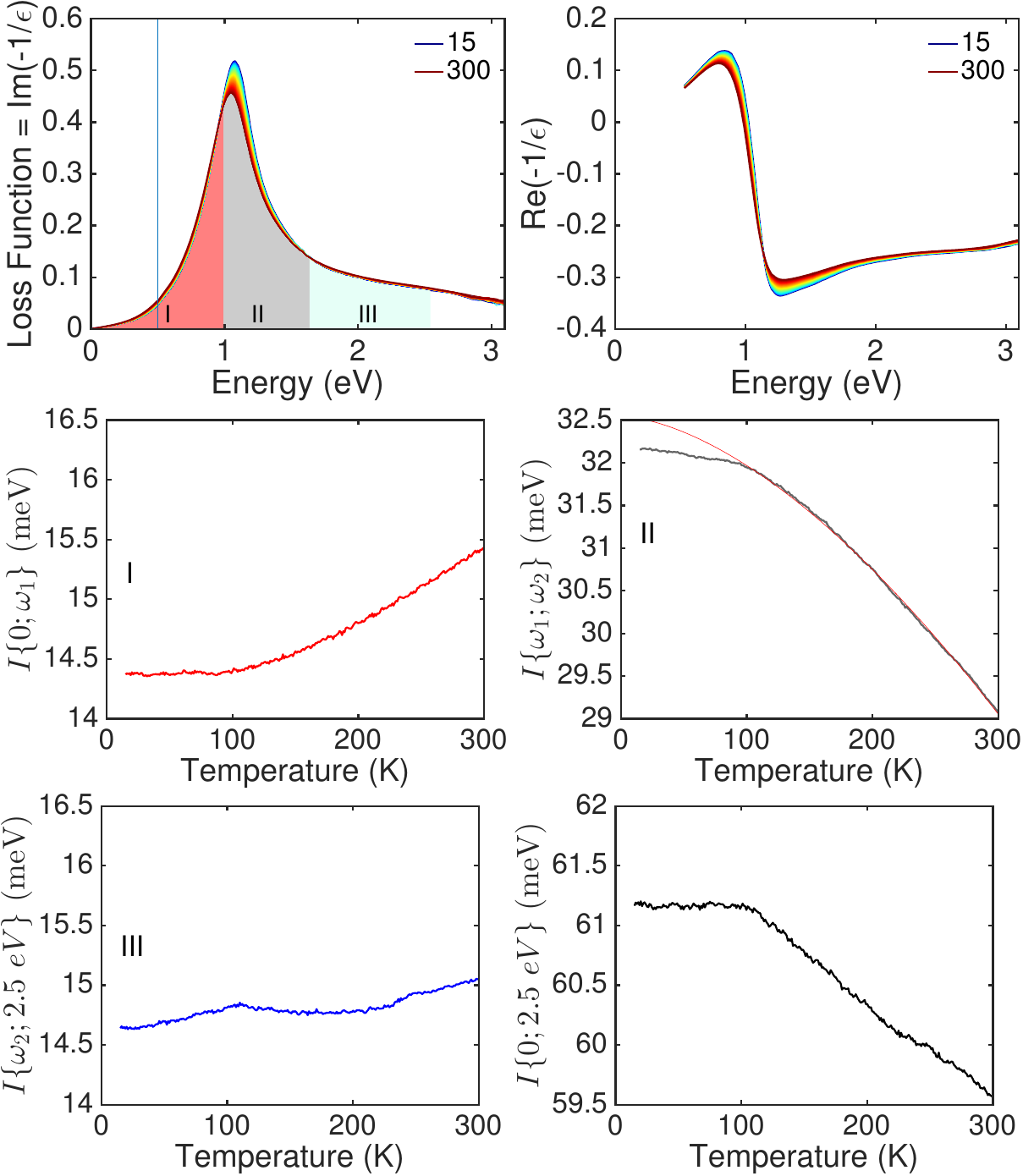}
\caption{Sample Bi2212-81-OD. 
Caption details as in Fig.~\ref{fig:Bi2212-60-UD}. 
The value of the fitted exponent $\alpha$ is $1.6$.}\label{fig:Bi2212-81-OD}
\end{center}
\end{figure}
\begin{figure}[b!!!]
\begin{center}
\includegraphics[width=1.00\columnwidth]{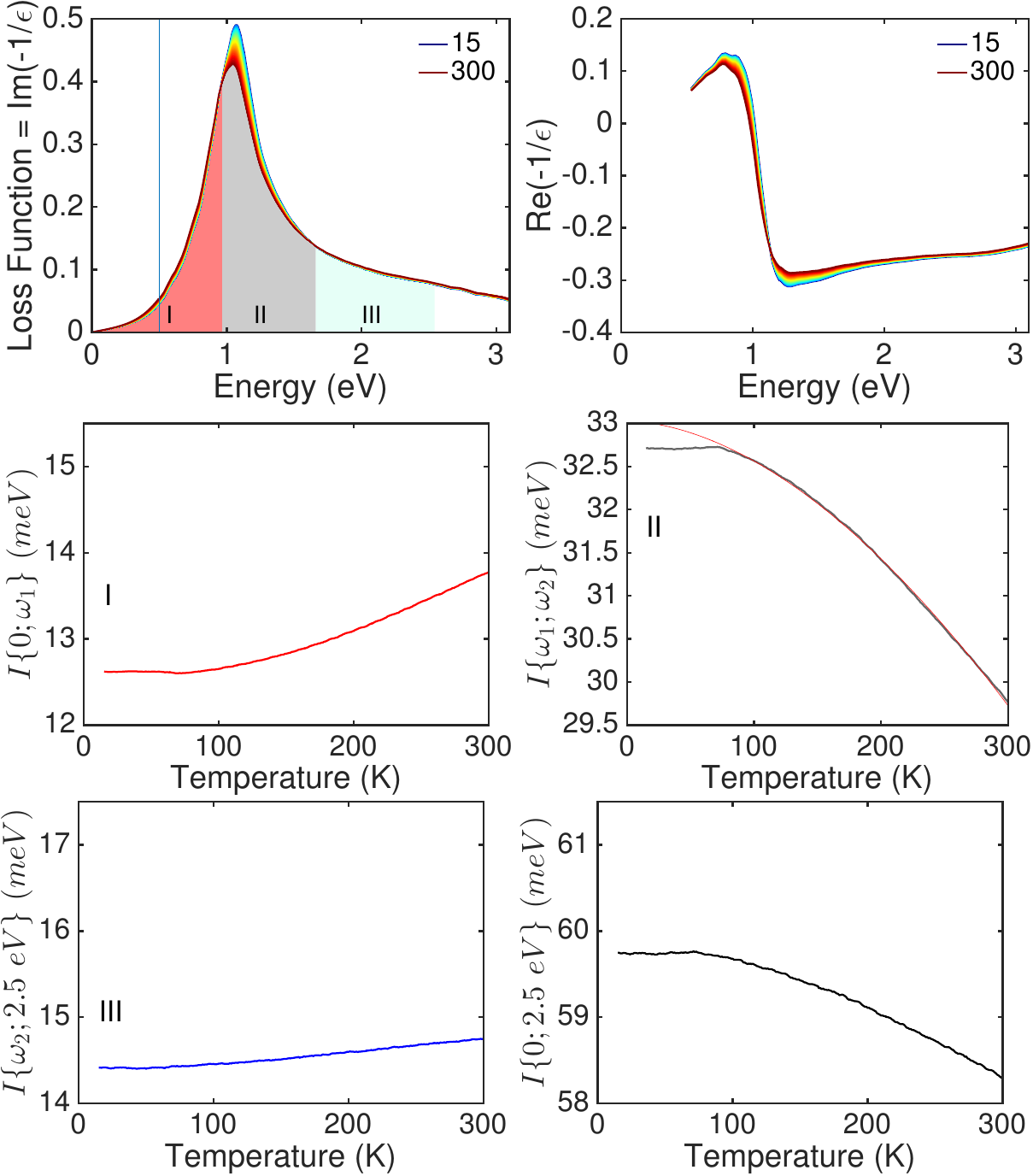}
\caption{Sample Bi2212-70-OD. 
Caption details as in Fig.~\ref{fig:Bi2212-60-UD}. 
The value of the fitted exponent $\alpha$ is $1.8$.}\label{fig:Bi2212-70-OD}
\end{center}
%
%
\begin{center}
\includegraphics[width=1.00\columnwidth]{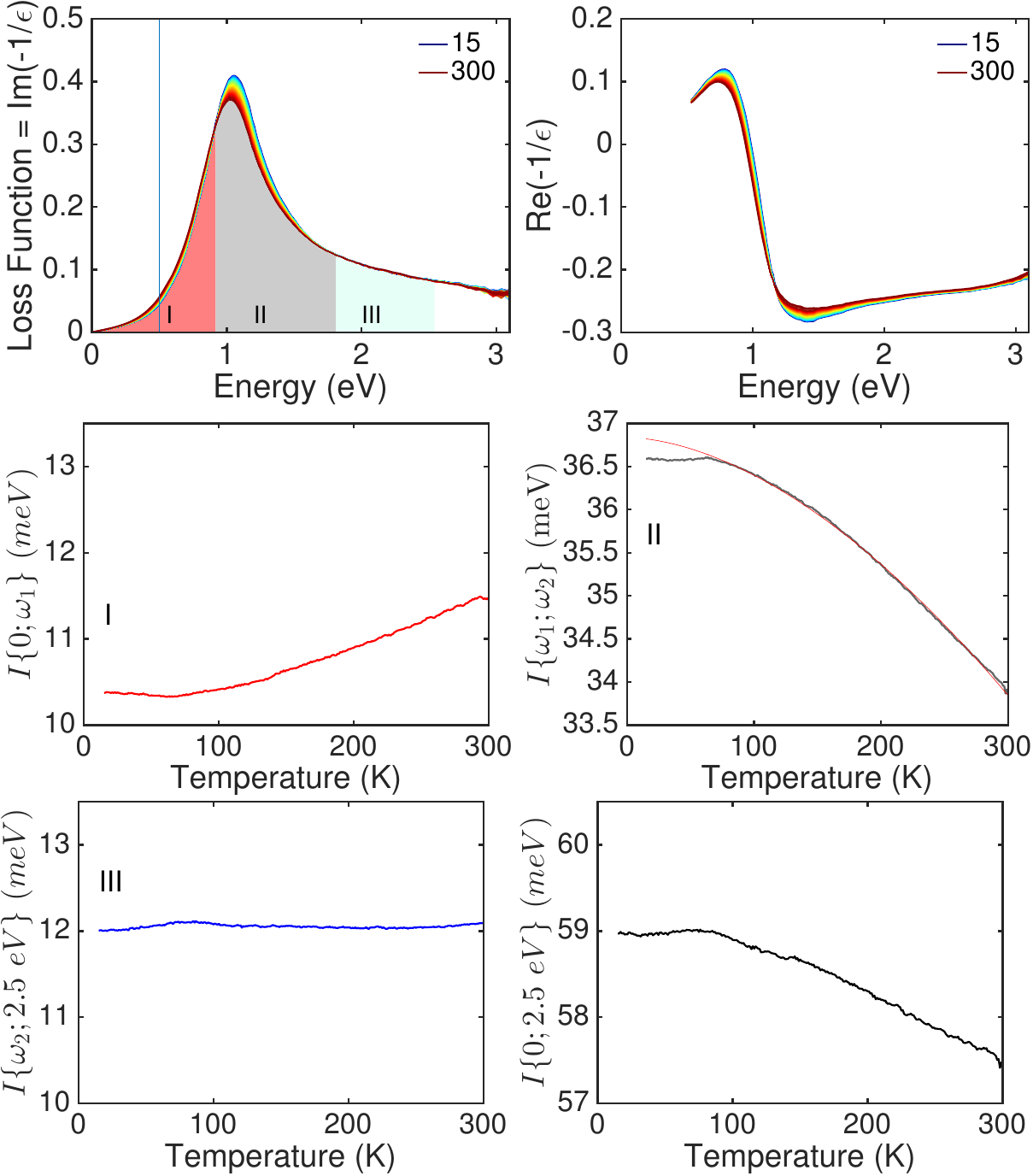}
\caption{Sample Bi2212-58-OD. 
Caption details as in Fig.~\ref{fig:Bi2212-60-UD}. 
The value of the fitted exponent $\alpha$ is $1.8$.}\label{fig:Bi2212-58-OD}
\end{center}
\end{figure}
\begin{figure}[t!!!]
\begin{center}
\includegraphics[width=1.00\columnwidth]{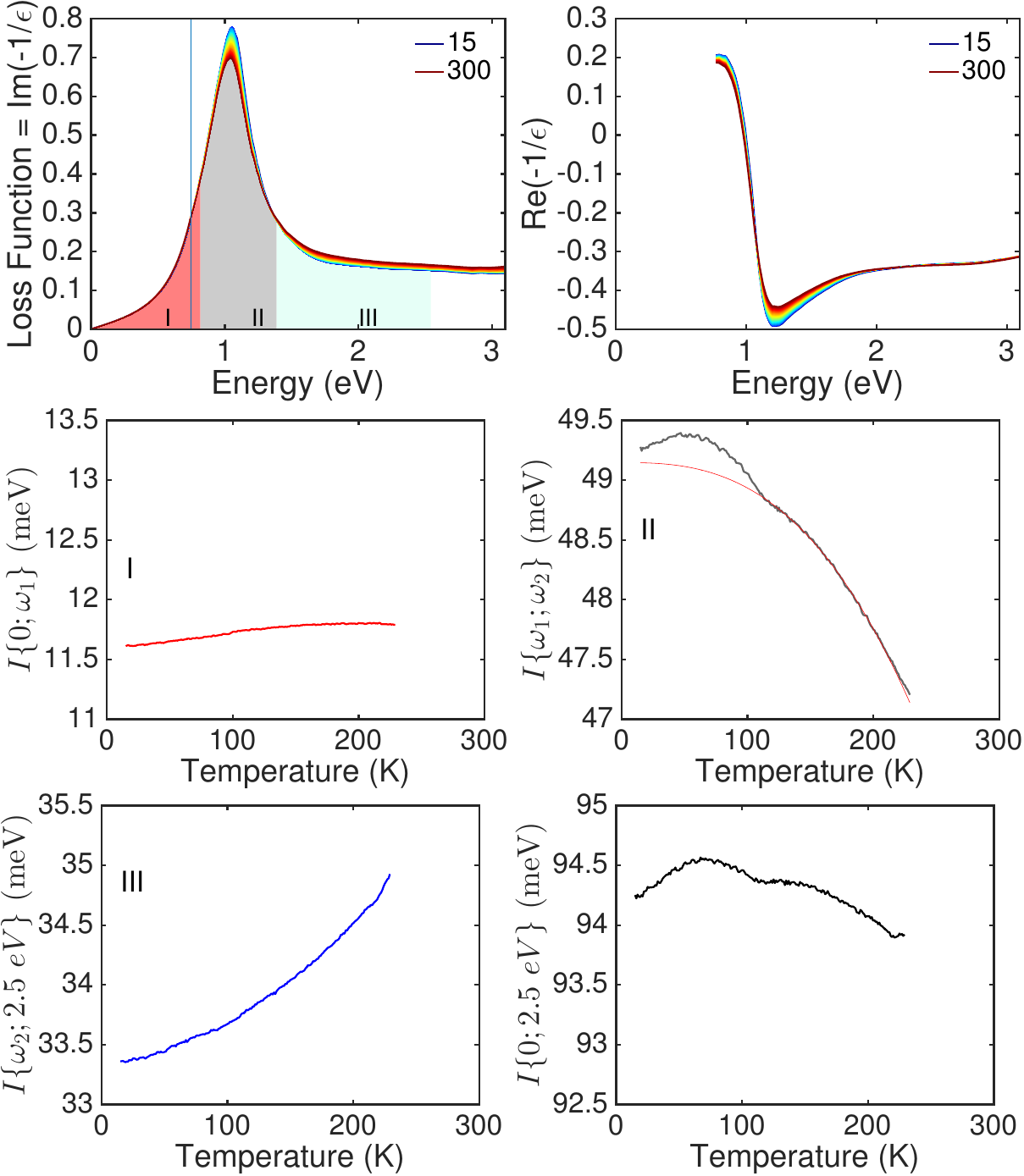}
\caption{Sample Bi2223-110-OP. 
Caption details as in Fig.~\ref{fig:Bi2212-60-UD}. 
The value of the fitted exponent $\alpha$ is 2.7. The evolution as a function of temperature above $220$~K was not fully reproducible due to a combination factors having to do with small crystal size and instrument drift. 
The temperature range for this sample is therefore limited below $220$~K.}\label{fig:Bi2223-110-OP}
\end{center}
\end{figure}
\clearpage
\clearpage
\section{Sample preparation}\label{appsample}
Optimally doped single crystals of Bi2223 were grown by the floating zone method. 
Due to the slow growth kinetics of the 3-layer compound, as compared to the 2-layer one, as well as the need of minimizing the formation of Bi2212 intergrowth in the Bi2223 crystals, dedicated growth conditions were chosen. 
The growth was performed at a very low travelling velocity ($\approx$ 50 $\mu$m/h) in a home-made 2-mirror furnace with a steep temperature gradient at the liquid-solid interface ($\approx$ 50$^\circ$C/mm) under a 7\%O$_2-93$\%Ar flowing atmosphere. 
The details of the growth of Bi2223 crystals are reported elsewhere~\cite{giannini2004}. 
The resulting as-grown crystals are slightly underdoped. In order to optimally dope and homogenize the oxygen content, cleaved crystals of typical $1-3$~mm size were annealed at 500$^\circ$C in 20 bar of O$_2$ for 50~h. 
As a result, the optimal $T_c = 110$~K and a transition width as narrow as $\le2$~K were obtained. 
The Bi2223 crystals used for the experiments described in this paper were selected out of dozens as being free of any 2212 traces in the X-ray diffraction pattern and any inflection at $\sim80$~K in the magnetic susceptibility. 
This indicates that the amount of Bi2212 intergrowth is well below 1 volume percent. 
Optimally doped Bi2212 single crystals with $T_c=91$~K were grown by using a floating zone method~\cite{wen2008}. 
Underdoped Bi2212 single crystals with a $T_c$ of 83~K were obtained by annealing in a sealed vacuum quartz tube tubes at 450$^\circ$C for three days. 
Underdoped Bi2212 single crystals with $T_c$'s of 70~K and 60~K were obtained by annealing during three days in a vacuum of 10$^{-2}$ Torr at 550$^\circ$C and 500$^\circ$C respectively.

\section{Ellipsometry measurements}\label{ellipso1}
In order to determine accurately the temperature and frequency dependence of the dielectric function {$\epsilon(\omega,q=0,T)$} in the infrared (from $0.5$~eV), visible and ultraviolet range, we performed spectroscopic ellipsometry using a commercial variable angle spectrometer (Woollam inc.). 
An ultrahigh vacuum cryostat (conflat flanges, no viton) of unique design allows continuous variation of the angle of incidence of the light, $\theta$ between 45 and 90 degrees with the surface normal, without breaking the vacuum. 
This is achieved by two arms composed of flexible metallic bellows terminated by optical windows that can be set parallel to the light path. 
Pitch, roll and yaw of the crystals are controlled with high precision. Sample temperature can be controlled from 10~K to 400~K. 
Samples are glued on a conical copper piece allowing rejection of light irrelevant to sample. 
A cold finger is thermally coupled with copper braids to the sample block, which is anchored mechanically to the bottom of the cryostat while remaining thermally isolated from it, insuring high mechanical stability upon temperature variation. 
Both a compact turbo-molecular pump and a compact ion-pump are mounted directly on top of the cryostat such as to keep pumping resistance to the minimum. 
After outgassing the cryostat walls by a heating-cooling cycle of several days, the valve to the turbo-pump is closed, the turbo is switched off, and the ion pump takes over to maintain the base pressure of $10^{-9}$~mbar. 
These precautions and the ultrahigh vacuum conditions are necessary requirements for stable sample surface conditions during the ellipsometric measurements at low temperature. 
Ellipsometry spectra are obtained during cool-down or warm-up in the temperature range from $15-300$~K at an average rate of 12~K per hour. 
The low noise level needed during the analysis of the present study required the best possible statistics and imposed the splitting of the spectral range in two: the first covers $0.5-1.5$~eV and the second covers $1.5-3.1$~eV. The angle of incidence was uniformly 70 degrees.
All data acquisition and temperature control is computer-controlled. 
\section{Presentation of the full data set}\label{fulldata}
In this section we present for each of the Bi2212 samples and the Bi2223 sample the loss function and the loss function integrals as a function of temperature in different frequency domains. 
In the top panels of Figs.~\ref{fig:Bi2212-60-UD}~--~\ref{fig:Bi2212-58-OD} are displayed: Imaginary (left) and real (right) part of $-\epsilon(\omega)^{-1}$ for selected temperatures. 
The curves in the interval on the left of the vertical blue line are Drude-Lorentz oscillator fits to the experimental complex loss function in the range $0.6-3.1$~eV.
Second and third rows from left to right: Integral of the loss function of the red, gray, blue and total area as a function of temperature. 
Right middle panel: Power law temperature dependence $B+AT^\alpha$, fitted to the data in the normal state.
\section{$c$-axis correction}\label{ellipso2}
\begin{figure}[h]
\begin{center}
\includegraphics[width=1\columnwidth]{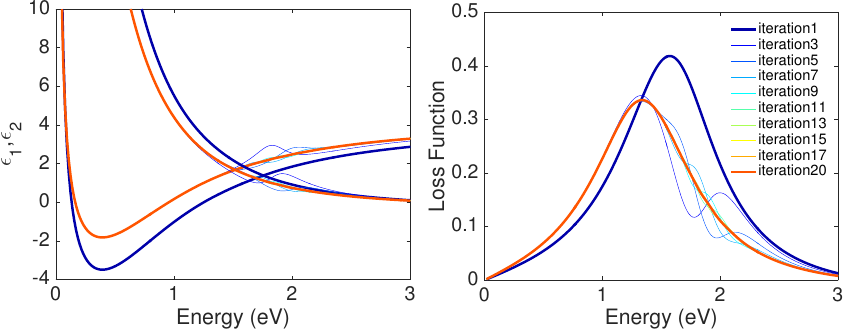}
\caption{Example of the $c$-axis correction obtained in 20 iterations: dielectric function (left) and corresponding loss function (right).} \label{Fig:iteration}
\end{center}
\end{figure}
Ellipsometry determines the ratio of $p$-polarized over $s$-polarized reflectivity coefficients, $\rho=r_p/r_s$, which in turn is a function of the tensor elements of the dielectric function and the angle of incidence relative to the surface normal, $\theta$. 
This relation is given by the Fresnel equations. In the interest of compactness of notation we define
\begin{equation}\label{alpha}
\eta\equiv \frac{\rho-1}{\rho+1}
\end{equation}
which from now on we treat as the primary experimental ellipsometric quantity. 
Here we consider the ellipsometric data for an optical uniaxial material with the optical axis perpendicular to the sample surface. 
The dielectric constant perpendicular to the sample surface is $\epsilon_c$, and along the sample surface we have $\epsilon_a$. 
The Fresnel equations give in this case
\begin{equation}\label{rho_aniso}
\eta=\frac{
(1+\delta-\epsilon_{a})\cos\theta\sqrt{\epsilon_a-\sin^2\theta}}
{\epsilon_{a}\cos^2\theta-(1+\delta)(\epsilon_{a}-\sin^2\theta)}
\end{equation}
where
\begin{equation}\label{smallparam}
\delta=\sqrt{\frac{1-\epsilon_{a}^{-1}\sin^{2}\theta}{1-\epsilon_c^{-1}\sin^{2}\theta}}-1
\end{equation}
The $\epsilon_a(\omega)$ spectra can be obtained from the combination of $\rho(\omega)$ and $\epsilon_c(\omega)$ using the method described below. 
In the isotropic case ($\epsilon_a=\epsilon_c\equiv \epsilon$) we have $\delta=0$ and Eq.~\eqref{rho_aniso} simplifies to
\begin{equation}\label{rho_iso}
\eta=\frac{\cos\theta\sqrt{\epsilon-\sin^2\theta}}{\sin^2\theta}
\end{equation}
Given the value of $\eta$ measured by ellipsometry, we can then calculate the dielectric function $\epsilon$ using the relation
\begin{equation}\label{aspnes0}
\epsilon=\sin^2\theta+\frac{\eta^2\sin^4\theta}{\cos^2\theta}
\end{equation}
We begin by noting that according to Aspnes~\cite{aspnes80}, even if the material is optically anisotropic, Eq.~\eqref{aspnes0} provides a good approximation of the tensor element of the dielectric constant along the intersection of the plane of reflection and the sample surface. 
For an uniaxial material with the optical axis perpendicular to the sample surface Eq.~\eqref{rho_aniso} can be solved iteratively, starting from Aspnes' zero'th order solution Eq.~\eqref{aspnes0}, {\em i.e.} $\epsilon_{a,0}=\epsilon$. 
The speed of convergence is controlled by the smallness of $\delta_j$ defined in Eq.~\eqref{smallparam} (with $\epsilon^{-1}_{a,j}$ instead of $\epsilon^{-1}_a$). 
The full solution of Eq.~\eqref{rho_aniso} is obtained by substituting $\epsilon_{a,0}$ in the right-hand side of the expression
\begin{equation}\label{iteration}
\epsilon_{a,j+1}=\sin^2\theta+\frac{\eta^2}{\cos^2\theta}
\left[\sin^2\theta-\frac{\delta_j\epsilon_{a,j}}{1-\epsilon_{a,j}+\delta_j}\right]^2
\end{equation}
The process continues by resubstituting $\epsilon_{a,j+1}$ in the right-hand side of the expression, which is reiterated until convergence is reached. 
Convergence takes typically less than 20 cycles, as illustrated by the example shown in Fig.~\ref{Fig:iteration}. 
The $c$-axis optical constants have been reported in Ref.~\onlinecite{tajima1993} (Bi2212) and Refs.~\onlinecite{petit2002,carbone2006b} (Bi2223), and have been found to be essentially independent of temperature for frequencies above 0.5 eV. Furthermore Re$\epsilon_c(\omega)\sim 3$-$5$ and Im$\epsilon_c(\omega)$ is very small. 
Due to crystal imperfections some $a$-axis admixture may have occurred in aforementioned experiments, due which the measured $\epsilon_c(\omega)$ may have been underestimated. 
Moreover, quite generally the bound-charge polarizability in the cuprates as obtained from $ab$-plane experiments corresponds to $\epsilon_{b} = 4.5\pm 0.5$ for Bi2212 and Bi2223. 
The anisotropy of the bound charge polarizability is known to be small in the cuprates, so that $\epsilon_{c}(\omega)$ (for which the free carrier contribution is negligible) should be near 4.5.
Anomalous spectral weight changes below $T_c$ in the $c$-axis response of underdoped cuprates has been discussed, reported and analyzed in~\cite{pwa95,Leggett1996,Tsvetkov1998,Basov1999,DvdM1996,Zelezny2001}. One may then wonder whether a $T$-dependence of the $c$-axis response, even a weak one, may interfere with the relatively small changes of the in-plane loss-function spectra that are reported here.
To investigate this possible influence, we analyze the case of optimally doped Bi2223. The same pseudo-dielectric function is fed in the procedure described above where this time $\epsilon_c(\omega, T)$ used in~\eqref{iteration} is measured by ellipsometry~\cite{carbone2006a}. The resulting loss function is shown in Fig.~\ref{Fig:Bi2223caxis}(a). Apart from a general level increase, there are no qualitative distinctions between this loss function and the one that we obtained with $\epsilon_c=3.5$ shown in Fig.~\ref{fig:Bi2223-110-OP}. The temperature-dependent loss function integrals in Fig.~\ref{Fig:Bi2223caxis}(b) yield extremely close temperature dependence and so do the corresponding Sommerfeld coefficients in Fig.~\ref{Fig:Bi2223caxis}(c) with matching characteristic temperatures. This has motivated us to adopt for the $c$-axis correction described above $\epsilon_c=4.5$ for Bi2212 and $\epsilon_c=3.5$ for Bi2223 throughout the frequency {{and temperature ranges} from $0.5-3.1$~eV.
\begin{figure}[h]
\begin{center}
\includegraphics[width=1\columnwidth]{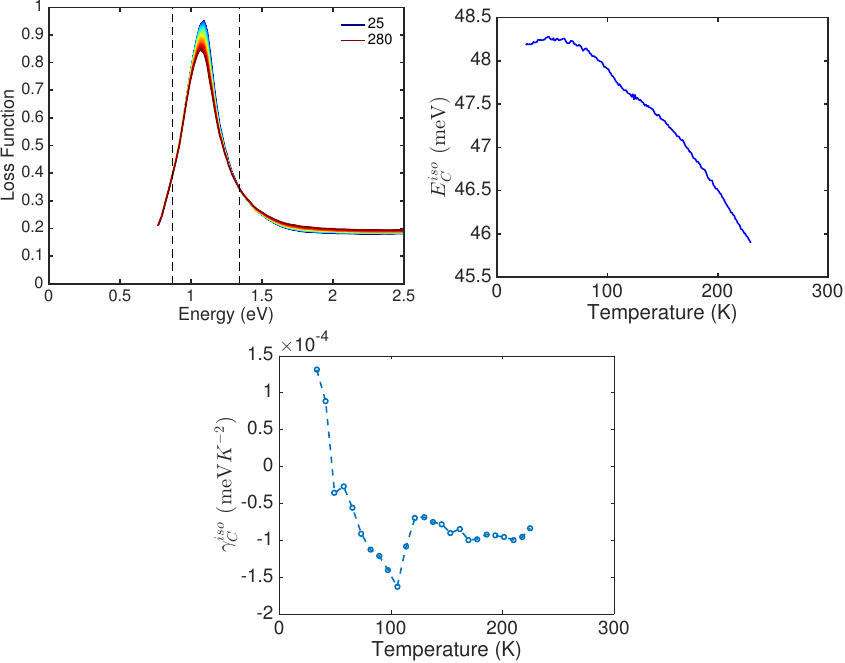}
\caption{Bi2223 $c$-axis correction using $\epsilon_c(\omega,T)$ from~\cite{carbone2006a}.} \label{Fig:Bi2223caxis}
\end{center}
\end{figure}
\section{Low energy Drude-Lorentz extrapolation}\label{DLExtrapol}
It was mentioned in section~\ref{Methods} that for a list of reasons, our analysis uses a spectral range limited by two isosbestics points, $\omega_1$ and $\omega_2$ displayed in Fig.~\ref{Fig:isopoints} for the under-doped Bi2212-83K sample. We show here that, in fact, the $0-\hbar\omega_1$~eV part of the loss function amounts to a marginal contribution to the partial Coulomb energy. To remain as close as possible to experimental input, we consider the Bi2223 loss function obtained with $c$-axis data correction to which the low energy part of the spectrum is complemented by far infrared data~\cite{carbone2006a} as shown in Fig~\ref{Fig:Bi2223Extrapol}(a). This region, situated below our measurement range, contains all previously reported superconductivity induced changes of temperature-dependent optical properties at photon energies, above the superconducting gap. However, these do not appear as strong features in the loss function at all: these spectral details are only distinguishable from a Drude-Lorentz fit (shown in Fig.~\ref{Fig:Bi2223Fit}) when displaying the data in a log-log scale, as shown in Fig~\ref{Fig:Bi2223Extrapol}(b). The temperature dependence of the loss function integral over the $0-\hbar\omega_1$ and the $0-2.5$~eV regions are shown in the second row of Fig.~\ref{Fig:Bi2223Extrapol}. It is obvious, when comparing the $0-2.5$~eV integral with the one carried-out on extrapolated data shown in Fig.~\ref{Fig:Bi2223Fit}, that aside an $\sim0.5$~meV offset, the temperature-dependence extracted from the two methods are identical. Additionally, one can also notice that the temperature dependence of the data at low energy is actually irrelevant as to the general behavior of the $0-2.5$~eV integral of the loss function.
\begin{figure}[h]
\begin{center}
\includegraphics[width=1\columnwidth]{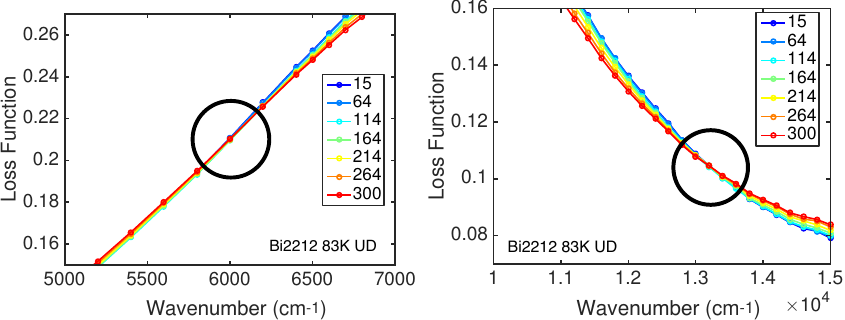}
\caption{Isosbestics points for Bi2212 83K UD} \label{Fig:isopoints}
\end{center}
\end{figure}
\begin{figure}[h]
\begin{center}
\includegraphics[width=1\columnwidth]{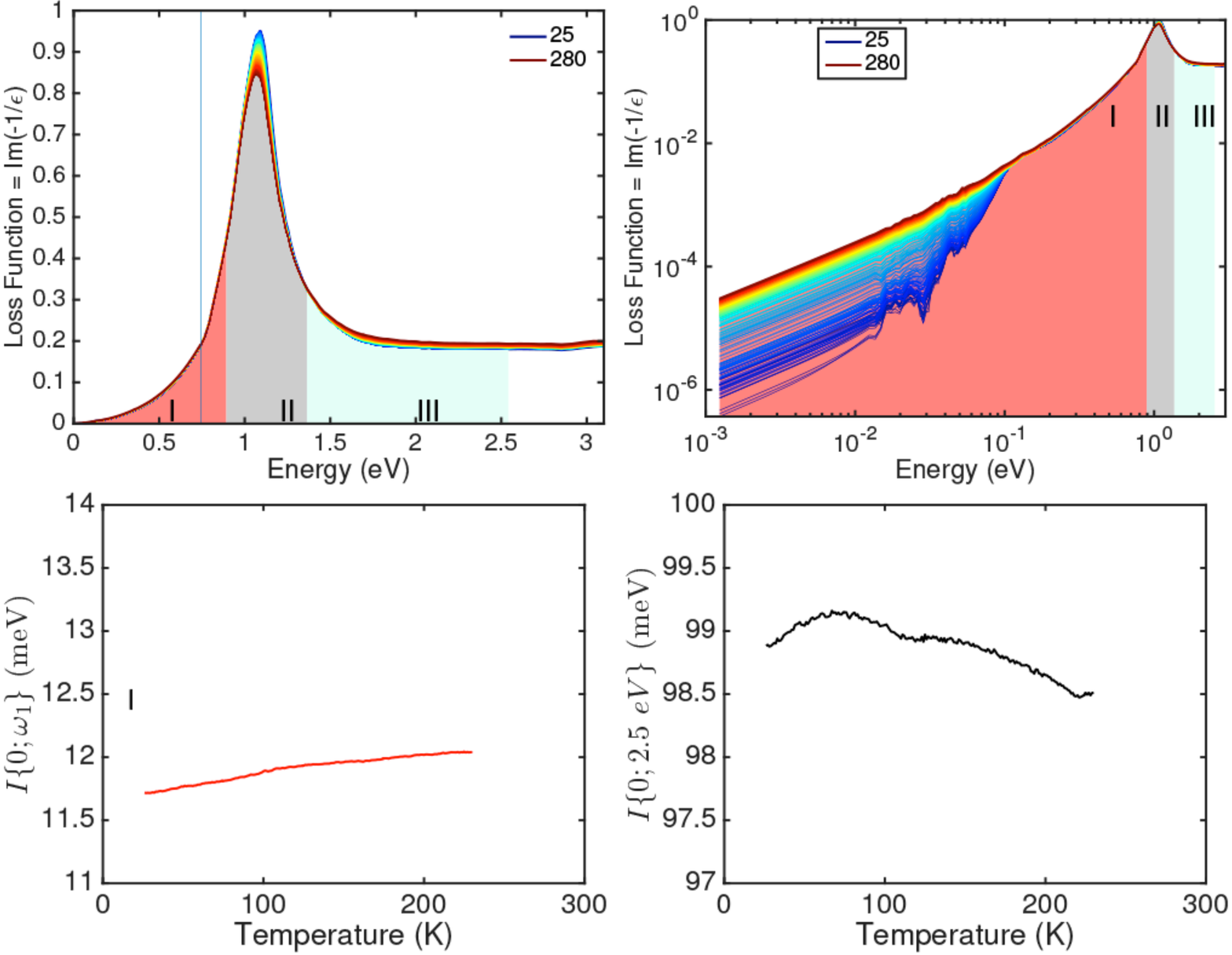}
\caption{Bi2223 optimally doped with extended low energy region from~\cite{carbone2006a}.} \label{Fig:Bi2223Extrapol}
\end{center}
\end{figure}
\begin{figure}[h]
\begin{center}
\includegraphics[width=1\columnwidth]{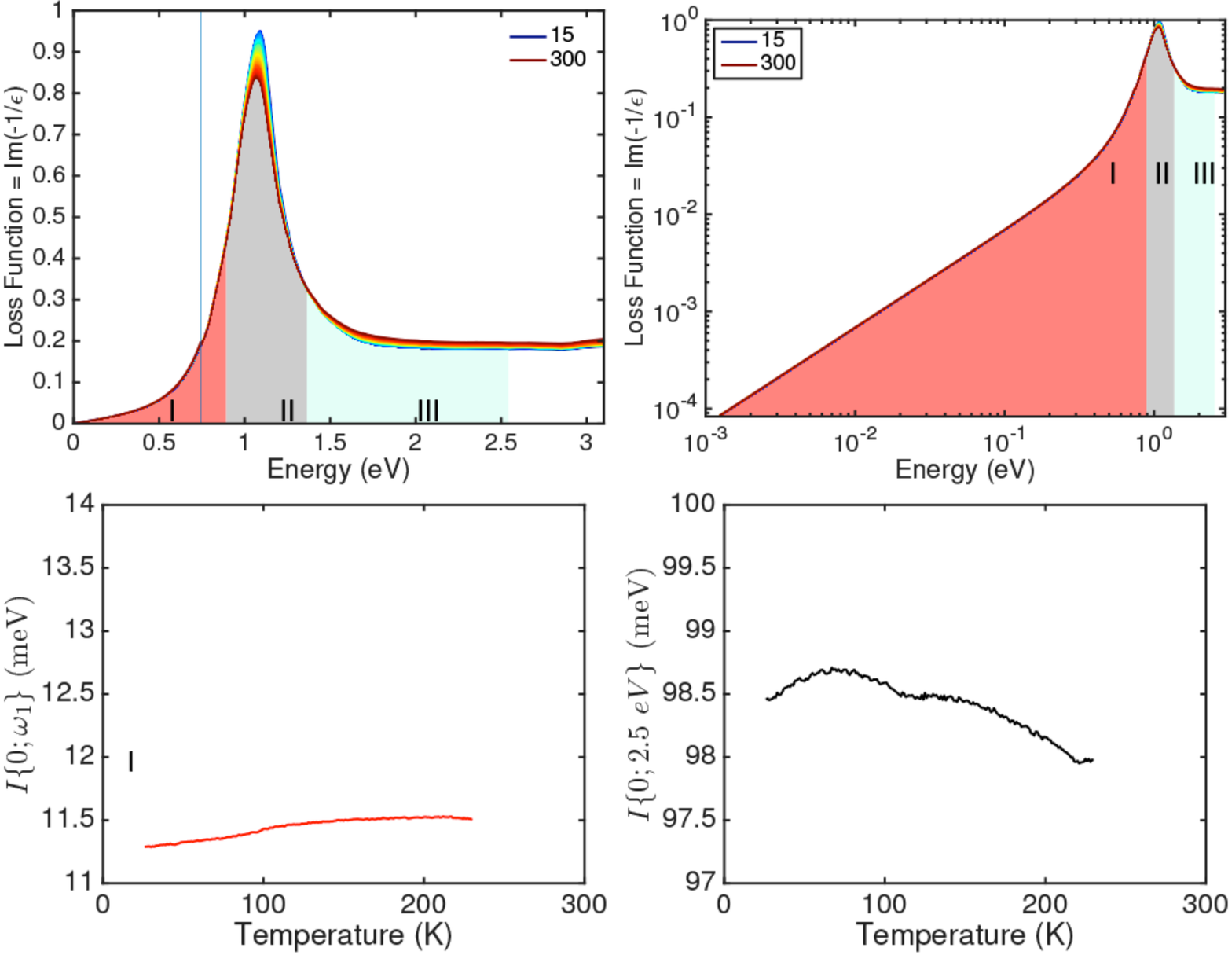}
\caption{Bi2223 optimally doped with low energy region extrapolated.} \label{Fig:Bi2223Fit}
\end{center}
\end{figure}
\section{Screening of the Coulomb interaction in a layered material}\label{Bound}
Our program in this paper is to use the data obtained in optical experiments, plus the ``extrapolation assumption'',  to infer something about the changes in the expectation value of the long- (but not very long-) wavelength part of the inter-conduction electron Coulomb interaction $E_C$. 
While in an isotropic 3D material this procedure is straightforward (see below), the strongly layered nature of the cuprates gives rise to a complication, as follows: As we will verify below, in a layered material with inter(multi)layer separations 
the screening of the Fourier component $V_q$ of the Coulomb interaction depends qualitatively on $q$: for $qs\ll1$ it is of the standard ``bulk 3D'' form and is completely taken into account by a $q$-independent dielectric constant $\epsilon(\omega)$, for $qs\gtrsim 1$ the effect is more complicated and for $qs\gg 1$ is represented by a $q$-dependent ``pseudo-dielectric constant'' $\epsilon_{ps}=1+[qs/2][\epsilon(\omega)-1]$. 
This is because, while in 3D the Fourier component is proportional to $q^{-2}$, in 2D it is $\propto q^{-1}$. 
Since in the MIR scenario the main contribution to $E_C$ comes from $qs \gtrsim 1$ with a strictly 2-dimensional $q$, but the optics measures the dielectric tensor for $qs\ll 1$ where $q$ is 3-dimensional, some care is necessary.

In the following we assume until further notice, as in Refs.~\onlinecite{Leggett1999a} and~\onlinecite{Leggett1999b}, (a) that any screening of the inter-Coulomb energy by the ionic cores is itself ``3-dimensional'' and hence may be represented by a $q$-and $\omega$-independent constant $\epsilon_{sc}$, and (b) that inter-multilayer (inter-unit-cell) tunneling is negligible and hence that in the frequency range of interest the experimentally observed ($q=0$) c-axis dielectric function  is some $\omega$-independent constant. 
The effect of relaxing these assumptions will be briefly discussed in section~\ref{C-axis_contribution}. For pedagogical simplicity we will give the explicit discussion for a single-plane cuprate; the generalization to the actual (bilayer) case of interest is straightforward and will be indicated where necessary. 

 The general statement following from the fluctuation-dissipation theorem is that at $T=0$ (which we adopt throughout this subsection), the full susceptibility $\chi(q,\omega)$ and the $\mathbf{q}$-th Fourier component of the Coulomb interaction are related as
\begin{equation}\label{AJL1a}
      E_C^{\mathbf{q}}=
      \frac{\hbar}{2\pi}\mbox{Im} \int_0^{\infty} 
      V_\mathbf{q}\chi(\mathbf{q},\omega) d\omega
\end{equation}     
where $V_\mathbf{q}$ is the Fourier transform of the bare Coulomb interaction. For a bulk 3D sample this implies a direct relation to the longitudinal dielectric function
\begin{equation}\label{AJL1b}
     E_C^{\mathbf{q}}=
      \frac{\hbar}{2\pi}\mbox{Im} \int_0^{\infty} d\omega
      \frac{-1}{\epsilon(\mathbf{q},\omega)}
\end{equation}     
However, ${\epsilon(\mathbf{q},\omega)}$ characterizes the dielectric response of a 3D material; consequently Eq. \eqref{AJL1b} can not be directly applied to, {\em e.g.}, the case of a two-dimensional conducting plane. The general formula irrespective of the geometry is given by 
\begin{equation}\label{AJL1}
      E_C^{\mathbf{q}}=
      \frac{\hbar}{2\pi}\mbox{Im} \int_0^{\infty} d\omega
      \frac{-1}{1+V_\mathbf{q}\chi^{(0)}(\mathbf{q},\omega)}
\end{equation}     
where $\chi^{(0)}(\mathbf{q},\omega)$  is defined by the relation
\begin{equation}
\chi(\mathbf{q},\omega)^{-1} = \chi^{(0)}(\mathbf{q},\omega)^{-1} + V_\mathbf{q},
\end{equation}
Note that this is a {\em definition} of $\chi^{(0)}(\mathbf{q},\omega)$, which despite resemblance to the random phase approximation, does not rely on the validity of that approximation. However, we shall assume for present purposes that in the $(\mathbf{q}, \omega)$ region of interest the $ab$-plane component ($q\neq 0$, $q_z=0$ where $q$ and $q_z$ are the $ab$-plane and $c$-axis components respectively),  that $V\chi^{(0)}$ is a function only of $\omega$ (this subsumes the ``extrapolation assumption'') and that for the $c$-axis component  ($q_z\neq 0$, $q=0$)  $V\chi^{(0)}=0$, and thus the $c$-axis component does not contribute to Eq. \eqref{AJL1}; we return to it in appendix~\ref{C-axis_contribution}.

The three-dimensional Fourier transform of the Coulomb potential in a layered electron gas is \cite{Fetter1974,Morawitz1993}
\begin{equation}\label{Vq}
     V_{\mathbf{q}}=\frac{e^2 s}{2\epsilon_0 q} \frac{\sinh{qs}}{\cosh{qs}-\cos{q_zs}} 
\end{equation}
In the limit $|\mathbf{q}|s\ll1$ this is just $e^2/(\epsilon_0 |\mathbf{q}|^2)$, and hence $E_C^{\mathbf{q}}$ is simply the integral over the familiar loss function Im$\epsilon(\mathbf{q},\omega)^{-1}$. 
However, the regime which dominantly contributes to the overall Coulomb energy is $qs\gtrsim 1$. If we make the approximation $qs\gg 1$, then for any $q_z$ $V_q$ is approximately given by  $e^2s/(2\epsilon_0 q)$, so after summation over $q_z$ we find the ``per-plane'' result 
\begin{equation}\label{Vq3}
    E_C^{\mathbf{q}}=  
     \frac{\hbar}{2\pi}\mbox{Im} \int_0^{\infty} d\omega     
     \frac{-1}{1+(qs/2)[\epsilon_a(\omega)-1]}          
\end{equation}
In section~\ref{C-axis_contribution} a more general formula (Eq.
\eqref{c-axis_corra}) 
for $E_C^{\mathbf{q}}$  is given, and it is shown that after summation over $q_z$ the result integrated over $q$ up to a cutoff $q_0$ coincides with \eqref{Vq3} provided $q_0 s\gg 1$. 

However, we wish to calculate the value of the {\em screened} Coulomb interaction between the conduction electrons. This can be done simply by replacing $V_q$ by $V^{sc}_q=V_q/\epsilon_{sc}$ where $\epsilon_{sc}$ is the frequency- and wave-vector-independent dielectric constant due to screening of the conduction electrons by the ionic cores  (recall we are assuming this screening to be 3-dimensional, {\em i.e.} due uniformly to the whole unit cell). 
In the limit $qs\ll1$ the effect is simply to multiply the expression \eqref{AJL1} by an overall factor of $\epsilon_{sc}$. 
We may see this more explicitly as follows: 
The general form of the Hamiltonian of a metal is (see Eq.~\eqref{hamiltonian})
\begin{equation}
\hat{H}=\hat{T}_{}+\hat{U}+ \hat{V}_{C}
\end{equation}
where $\hat{T}$ is the kinetic energy, $\hat{U}$ the crystal potential, and
\begin{equation}
\hat{V}_{C}=\frac{1}{2}\sum_\mathbf{q}V_\mathbf{q}\hat{\rho}_\mathbf{q}\hat{\rho}_{-\mathbf{q}}
\end{equation}
The dielectric function of such a many-electron system is
\begin{equation}
\epsilon(\mathbf{q},\omega)=1+V_\mathbf{q}\chi^{(0)}(\mathbf{q},\omega)
\end{equation}

Often one is interested in the free carrier properties, in which case it is useful to integrate out the degrees of freedom having to do with the ``bound charge'', {\em  i.e.} the interband transitions. This distinction is meaningful when the interband transitions are well separated from the intra-band degrees of freedom, which is actually the case in the cuprates. The dielectric constant can be split as follows
\begin{equation}\label{G6}
\epsilon(\mathbf{q},\omega)=1+S(\mathbf{q},\omega)+V_\mathbf{q}\chi^{(0)}(\mathbf{q},\omega)
\end{equation}
where $S(\mathbf{q},\omega)$ subsumes all bound charge terms. At low frequencies and small $q$ we have
\begin{equation}\label{G7}
1+S(\mathbf{q},\omega)=\epsilon_{sc}
\end{equation}
In the sequel we will neglect $q$ and $\omega$ dependencies up to frequencies $\omega_b$ which is some high energy scale representative of the interband transitions. The dielectric function at low frequencies becomes
\begin{equation}\label{eq:epslow}
{\epsilon}(\mathbf{q},\omega)=\epsilon_{sc}\left\{1+{V}_\mathbf{q}^{sc}\chi^{(0)}(\mathbf{q},\omega)\right\}
\end{equation}
where ${V}_\mathbf{q}^{sc}=V_\mathbf{q}/\epsilon_{sc}$. The effective low energy Hamiltonian is
\begin{equation}
 \hat{{H}}^{eff}=\hat{K} +\hat{{V}}^{sc}_C
\end{equation}
where the first term describes the free charge carrier band dispersion and the second one the bare Coulomb interaction screened by aforementioned ``bound charges'', so that 
\begin{equation}
\hat{{V}}^{sc}_C=\frac{1}{2}\sum_\mathbf{q} {V}_\mathbf{q}^{sc}\hat{\rho}^{f}_\mathbf{q}\hat{\rho}^{f}_{-\mathbf{q}}
\end{equation}
Consequently the relation between low energy susceptibility taking into account the screened interaction $\chi^{sc}(\mathbf{q},\omega)$, and $ \chi^{(0)}(\mathbf{q},\omega)$ becomes
\begin{equation}
\chi^{sc}(\mathbf{q},\omega)^{-1} = \chi^{(0)}(\mathbf{q},\omega)^{-1} +V^{sc}_{\mathbf{q}}
\end{equation}

The fluctuation dissipation theorem~\cite{Nozieres1959} provides the relation between the screened Coulomb interaction and the limited range susceptibility integral
\begin{equation}\label{eq:chi}
\frac{\hbar}{\pi} \int_0^{\omega_b} \mbox{Im}\chi^{sc}(\mathbf{q},\omega)d\omega
=
\langle \hat{\rho}^{f}_\mathbf{q}\hat{\rho}^{f}_{-\mathbf{q}} \rangle 
\end{equation}
from which, with the help of Eqs.~\eqref{eq:epslow} and~\eqref{eq:chi}
\begin{equation}\label{eq:esc}
{E}^{mir}_C=\langle\hat{{V}}^{sc}_C\rangle=
\frac{\hbar\epsilon_{sc}}{2\pi}\sum_\mathbf{q} \int_0^{\omega_b} \mbox{Im} \frac{-1}{{\epsilon}(\mathbf{q},\omega)}d\omega 
\end{equation}
Eq.~\eqref{eq:esc} tells that the loss function integral, when carried out over the free-carrier part of the response, probes a fraction $1/\epsilon_{sc}$ of the Coulomb interaction between the free charge carriers, which --~compared to the bare Coulomb interaction~-- is already reduced by an additional factor $1/\epsilon_{sc}$. 
To verify that no double counting of $1/\epsilon_{sc}$ has occurred, we reformulate the individual $q$-terms of LHS and RHS of Eq.~\eqref{eq:esc}
\begin{equation}
\frac{\hbar}{2\pi} \int_0^{\omega_b} \frac{{V}_\mathbf{q}^{sc} \mbox{Im}\chi^{(0)}(\mathbf{q},\omega)}{\left|1+{V}_\mathbf{q}^{sc}\chi^{(0)}(\mathbf{q},\omega)\right|^2}d\omega
=
\frac{1}{2}\langle {V}_\mathbf{q}^{sc}\hat{\rho}^{f}_\mathbf{q}\hat{\rho}^{f}_{-\mathbf{q}} \rangle 
\end{equation}
which in the weak coupling limit ($V_\mathbf{q} \rightarrow 0$, so that the denominator $\rightarrow 1$) returns the fluctuation-dissipation theorem for the non-interacting system
\begin{equation}
\frac{\hbar}{\pi}\int_0^{\omega_b} \mbox{Im}\chi^{(0)}(\mathbf{q},\omega)d\omega
=\langle \hat{\rho}^{f}_\mathbf{q}\hat{\rho}^{f}_{-\mathbf{q}} \rangle
\end{equation}
as expected. The remaining fraction ($1-1/\epsilon_{sc}$) of the screened interaction energy is recovered in the loss-function in the range of interband transitions. Since the interband region is typically smeared out over several tenths of eV, the corresponding signatures are small and very difficult to detect experimentally.

 In the more relevant case of a layered system with  $qs\gg 1$ we can go through the same argument, but must now bear in mind that $V^{sc}_q\chi(q,\omega)$ is no longer equal to $[\epsilon(\omega)/\epsilon_{sc}-1]$ but rather to $[qs\epsilon(\omega)/\epsilon_{sc}-1]$. Thus, we recover Eqs. (4.1.4) and (4.1.5) of Ref.~\onlinecite{Leggett1999b}, and hence Eq. \eqref{eq14} of the main text.
Finally, if we relax the assumption that the core screening is ``three-dimensional'' (for example, assume that it comes only from the highly planar array of intra-layer Cu's and O's) the effect is simply to replace the quantity $\epsilon_{sc}$ in Eq.~\eqref{eq14} by the relevant $\epsilon_{sc}(q)$, which may have a substantial $q$-dependence, but in practice makes not much difference to the computed value of ${E}^{mir}_C$.

\section{C-axis contribution in a layered material}\label{C-axis_contribution}
For the $c$-axis the free carrier spectral weight is very low. The transition from normal to superconducting state is characterized by the appearance of a Josephson plasmon. 
For the Bi2212 materials the Josephson plasma frequency is much lower than the (already low) frequency observed in single-layer Tl2201~\cite{Tsvetkov1998}, in fact the exact value is not known since it is below the experimental window of infrared spectroscopy. However, Zelezny et al.~\cite{Zelezny2001} observed a temperature dependence in the optical phonon range, which they attributed to the transverse optical Josephson plasmon~\cite{DvdM1996}. In order to have an influence on the ellipsometric data in the range of the $ab$-plane plasmon, it would be necessary that the optical spectral weight associated to such a plasmon is transferred from high energy. There are no indications for this; in fact the absence of reflectivity changes in the near infrared in these compounds (as discussed for example in Appendix~\ref{ellipso2}) rather suggests that this spectral weight is reshuffled within the infrared range. If we nonetheless model the $c$-axis dielectric function with a 400 cm$^{-1}$ plasma frequency which disappears in the normal phase (equivalent to a transfer of the associated optical spectral weight to infinite frequency), we obtain a strongly overestimated upper bound of the $c$-axis contribution to the Coulomb energy. To simulate the effect on the optical properties we use the following expressions
\begin{align}
\epsilon_c(\omega,T)&=\epsilon_{sc}-\frac{\omega_{pc}^2 n_s(T)}{\omega^2} &\label{contribepsilonc}\\
n_s(T)&= \begin{cases} 1-(T/T_c)^2, & T\leq T_c \\ 
0, & T>T_c \end{cases} \nonumber
\end{align}
with the parameters $\omega_{pc}/2\pi c= 400$~cm$^{-1}$ and $\epsilon_{sc} = 4.5$. 
We furthermore assume, as per MIR scenario, that the value of the loss-function can be extrapolated up to an upper limit $q_0$, and for larger $q$ has negligible contribution to the S--N difference of the Coulomb energy. 
This yields the $q$-averaged {\em effective} Coulomb energy in the MIR scenario (see Eqs. \eqref{eq:esc} and \eqref{AJL1b}) 
\begin{equation}\label{L_av0}
{E}^{mir}_C=
\frac{\hbar\epsilon_{sc}a^2s}{8\pi^3 }\mbox{Im} 
\int_0^{\omega_b}  d\omega 
\int_0^{q_0}  qdq \int_{-\pi}^{\pi}dq_z  \frac{-1}{{\epsilon}(\mathbf{q},\omega)} 
\end{equation}
where the upper bound $\omega_b$ is the energy scale of the core-electron excitation energies (see section~\ref{Bound}), and
\begin{equation}
\epsilon(\mathbf{q},\omega)=1+{V}_{\mathbf{q}}\chi^{(0)}(\mathbf{q},\omega)
\end{equation}
with $\chi_0(\mathbf{q},\omega)$ defined in Eq. (4.1.2) of Ref.~\onlinecite{Leggett1999b}.  
Without loss of generality we can define the tensor $K_{i,j}(\mathbf{q},\omega)$ in the following way
\begin{equation}
  \chi^{(0)}(\mathbf{q},\omega) \equiv \frac{2\epsilon_0} {e^2s} \sum_{i,j} K_{i,j}(\mathbf{q},\omega)q_iq_j
\end{equation}
In a 2D system this becomes simply 
\begin{equation}\label{fetter_eps}
 \chi^{(0)}(q,\omega) =\frac{2\epsilon_0} {e^2s}  K(q,\omega)q^2
\end{equation}
where $K_{}(q,\omega)$ is the in-plane component~\cite{Leggett1999b} here assumed to be isotropic within the plane.
Allowing for finite tunneling between the layers does not change $V_q$ (given by Eq.  Eq.~\eqref{Vq}) as long as we stick to the model of the $\delta$-layers, but it introduces a non-zero $k_z$ dispersion in the single particle dispersion with the property $\epsilon_{k,k_z+2\pi/s}=\epsilon_{k,k_z}$; consequently $\chi^{(0)}$ should be a $2\pi/s$ periodic function of $q_z$.
Since $\chi^{(0)}$ should vanish as $q^2+q_z^2$ for small momentum, it is described by a series in powers of $\tilde{q}_z^{2n}$, with $\tilde{q}_zs/2=\sin(q_zs/2)$ and  $n$ an integer number. To avoid clutter in the evaluation of the $q_z$ integral later in this appendix -but admittedly at the cost of loss of generality- we truncate this series at $n=1$, and obtain
\begin{eqnarray}
\chi^{(0)}(q,q_z,\omega)=
\frac{2\epsilon_0} {e^2s}\left[K_{}(q,\omega)q^2
+
K_{z}(q,\omega)\tilde{q}_z^2\right]
\end{eqnarray}
We furthermore use the MIR Ansatz that $\epsilon(q,\omega)$ has no important dispersion at least up to $q_0$, so that we can remove the $q$ dependence of $K_{}$  and $K_{z}$.
In parallel to the free carrier response there exists a bound-charge screening described by the function $S(\mathbf{q})$ which can be considered static in the range of the plasma frequency. The corresponding  bound-charge dielectric function is $\epsilon_{sc}(\mathbf{q})=1+S(\mathbf{q})$, so that 
\begin{equation}\label{epsilon5}
\epsilon(\mathbf{q},\omega)=\epsilon_{sc}(\mathbf{q})+
\frac{\sinh{qs}}{q}\frac{q^2K_{}(\omega)+\tilde{q}_z^2K_{z}(\omega)}{\cosh{qs}-\cos{q_zs}} 
\end{equation}
\begin{figure}[h!!!]
\begin{center}
\includegraphics[width=\columnwidth]{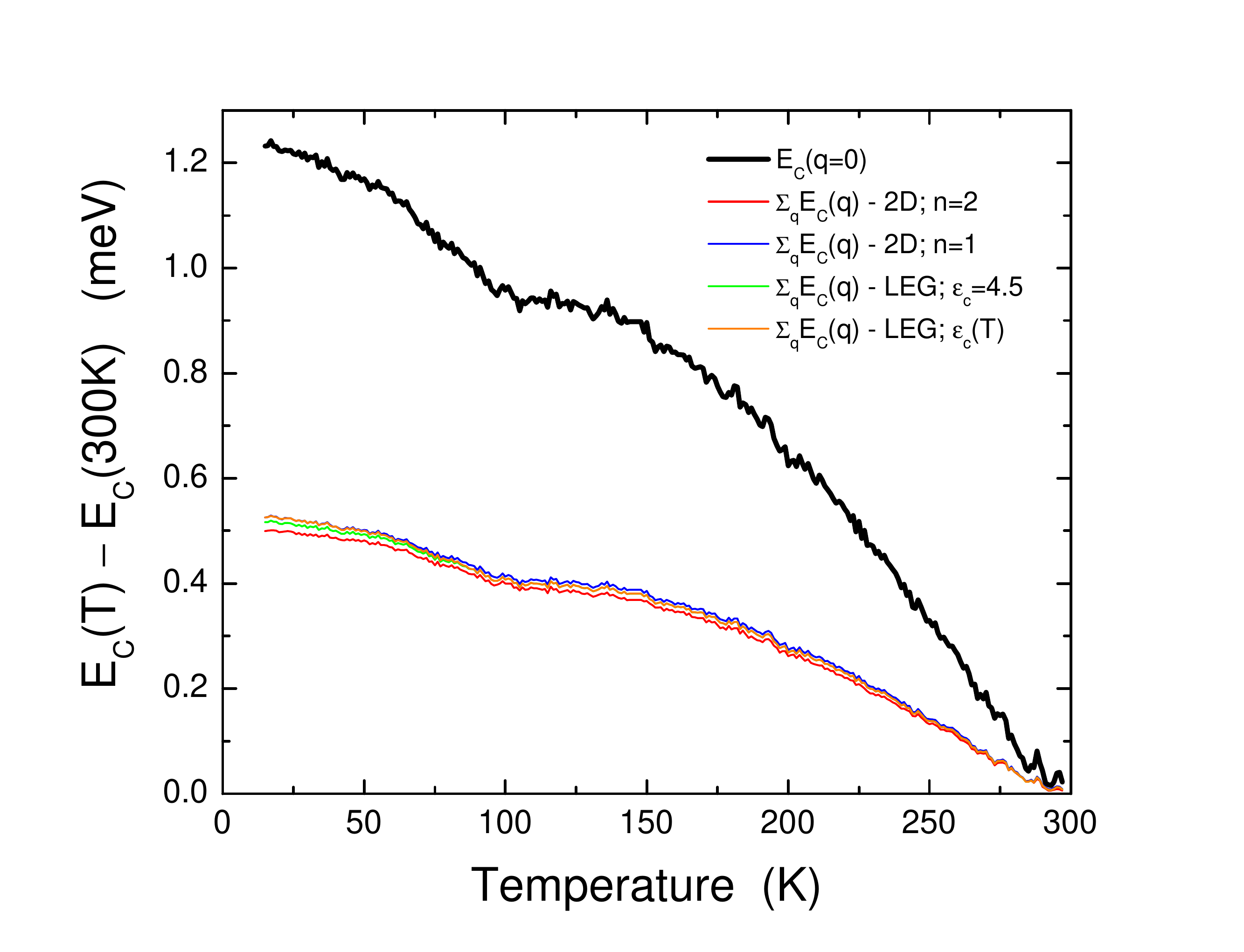}
\caption{\color{black}Relative change of the integral of the Loss function between $0$ and $2.5$~eV with respect to room temperature of sample Bi2212-91-OpD. 
Black: the ``$q=0$'' result, same as in the lower right of Fig.~\ref{fig:Bi2212-91-OP}. 
Red, blue, green and orange curves are the finite momentum extrapolation to finite $q_0=0.31$~\AA$^{-1}$, using different extrapolation schemes:
Red: 2D bilayer expression, Eq.~\eqref{eq14} with $n=2$, and $d=3.2$~\AA. 
Blue: 2D monolayer expression, Eq.~\eqref{Lav3}. 
Green: layered electron gas (LEG) expression, Eq.~\eqref{c-axis_corra} with  $\epsilon_c=\epsilon_{sc}=4.5$ 
Orange:  LEG expression, Eq.~\eqref{c-axis_corra} with $\epsilon_c(T)$ described by Eq.~\eqref{contribepsilonc}. 
The overlap of orange and green curves indicates that the c-axis loss function has negligible influence on the temperature dependence of the Coulomb energy. The overlap of red, blue and green curves implies that the extrapolation has negligible dependence on the type of extrapolation scheme chosen.
While the black curve is the expression for the "true" Coulomb energy which one would get by simply multiplying the optically measured loss function by the area of the first Brillouin zone, the colored curves are expressions for the "MIR Coulomb energy" as discussed in section~\ref{section4}. The fact that the colored curves are a substantial fraction of the black one -rather than being related to it by approximately the ratio of the disk $q<0.31A^{-1}$ to that of the first Brillouin zone- is largely attributable to the replacement of $E_C$ by ${E}^{mir}_C$, see Eqs. \eqref{eq:esc} for the 3D case and \eqref{Lav3} for the 2D case. 
}
\label{fig:caxis}
\end{center}
\end{figure}
If the momentum dispersion of $\epsilon_{sc}(\mathbf{q})$ is not too large, the dispersion has no important consequences for the Coulomb energy estimate; we will therefore simply replace $\epsilon_{sc}(\mathbf{q})$ by the constant $\epsilon_{sc}$ in the sequel of this appendix.
It is convenient at this point to introduce the following shorthand notation for the free carrier response parallel (perpendicular) to the conducting layers
\begin{eqnarray}\label{epsilon5'}
2K(\omega)/\epsilon_{sc}=fs\hspace{5mm}
2K_{z}(\omega)/\epsilon_{sc}=gs\hspace{5mm}
\end{eqnarray}
We insert the expression of the dielectric function, Eq. \eqref{epsilon5}, in the one for the Coulomb energy, Eq.~\eqref{L_av0}, and obtain after integration over $q_z$ 
\begin{align}\label{c-axis_corra}
&{E}^{mir}_C=\frac{\hbar a^2}{4\pi^2 s^2} \mbox{Im} 
\int_0^{\omega_b}d\omega   
\int _0^{q_0s} dx  \frac{x}{1+g_{}}  \\
&\left[-1+
\frac{
x^2f_{}/2-g_{}[x\coth x-1]
}
{\sqrt{
(x\coth x + x^2f_{}/2+g_{})^2-(1+g_{})^2x^2/\sinh^2x
}
}
\right]\nonumber
\end{align}
In the absence of  interlayer tunneling ($g_{}=0$) this reduces to Eq. (5.43) of Ref.~\onlinecite{presura2003}
\begin{align}\label{c-axis_corr}
{E}^{mir}_C=&\frac{\hbar a^2}{4\pi^2 s^2} \mbox{Im} 
\int_0^{\omega_b}d\omega   \\
&\int _0^{q_0}  dq 
\frac{q^2sf_{}/2}
{\sqrt{1+qsf_{}\coth qs + (qsf_{}/2)^2}}
\nonumber
\end{align}
In Ref.~\onlinecite{Leggett1999b} the case was discussed of a purely two-dimensional system of electrons,
where the effective screening of the inter-conduction electron Coulomb interaction is 3-dimensional and therefore described by the dielectric constant $\epsilon_{sc}$. This limit is described by $s\rightarrow\infty$, so that $\coth{qs}\rightarrow 1$. We rearrange the integrand to $1-(1+qsf_{}/2)^{-1}$, and substitute the definition~\cite{Leggett1999b} $sf_{}/2\equiv K(\omega)/\epsilon_{sc}$, with the result
\begin{equation}\label{Lav3}
{E}^{mir}_C = 
-\frac{\hbar a^2}{4\pi^2} \mbox{Im}\int_0^{\omega_b} d\omega
\int _0^{q_0}dq
\frac{q}{1+qK(\omega)/\epsilon_{sc}}
\end{equation}
where we recognize Eq. (4.1.5) of Ref. \onlinecite{Leggett1999b}. 

The result of Eq. \eqref{Lav3} is shown in Fig.~\ref{fig:caxis} and compared with  the extrapolation schemes of Eqs.~\eqref{c-axis_corra}, and ~\eqref{eq14} using $s=\bar{d}=7.8$~\AA, $q_0=0.31$~\AA$^{-1}$, and in Eq.~\eqref{eq14} $d=3.2\AA$, as well as the the integral of the experimental loss function for $q\approx 0$. From this comparison we conclude that the replacement of the ``true'' Coulomb energy by the MIR one, plus the extrapolation to finite momentum and the 0.31 A$^{-1}$ cutoff, effectively boils down to a uniform scaling of the $q=0$ result. The result is rather insensitive to the extrapolation scheme chosen, and corresponds to a scaling factor $F =0.42$. 

To determine the impact of temperature dependence of the $c$-axis dielectric function we compare the output of Eq.~\eqref{c-axis_corra} assuming that $\epsilon_c$ is independent of temperature (green curve), and that the temperature dependence is described by Eq.~\eqref{contribepsilonc} (orange curve). Despite somewhat exaggerated assumptions about the $c$-axis temperature dependence, the influence on the Coulomb energy is negligible in comparison to that of the $ab$-plane contribution. 
\section{Normal state $T$-dependence of ${E}_{C}^{iso}$}\label{Normaldata}
Quite generally increasing the quasi-particle relaxation broadens the loss function peak, and it causes a red-shift. 
The loss function integral will then exhibit a corresponding decrease in intensity. 
It is therefore of interest to see, if a relation exists between the temperature dependences of the loss function integral and the quasi-particle relaxation rate. 
Based on a similar reasoning it was demonstrated in Ref.~\onlinecite{norman2007} that the truncation at some finite value $\Omega$ of the spectral weight integral of the optical conductivity, Eq.~\eqref{K-sumrule}, introduces a temperature dependence of the quasi-particle relaxation rate. 
If indeed such a relation could be established, it would imply that in the relevant frequency range of about 1~eV, the relaxation rate at these frequencies would have the same temperature dependence for all dopings. 
At first glance this appears at odds with the fact that the transport relaxation rate is known to have strong qualitatively different behavior for different doping levels. 
However, we can not exclude a priori that the temperature dependence of $\gamma(\omega,T)$ is more universal among the cuprates for $\omega\sim 1~$eV. 
Additional intensity develops when the temperature passes through the zone between $T_{n2}$ and $T_{n1}$, where $T_{n2}$ is the lower bound of the region of $T^2$ temperature dependence, and $T_{n1}$ is the inflection point. 
Aforementioned additional intensity of the integrated loss function corresponds to a gain of the partial Coulomb energy for small $q$.
For the description of the dielectric properties of the interacting electrons in the normal state we adopt the generalized Drude model
\begin{equation}
\epsilon(\omega) = \epsilon_\infty- \frac{\omega_p^2/\omega}{\omega\left[1+\lambda(\omega)\right]+i\gamma(\omega)}
\label{drude}
\end{equation}
We furthermore use the model of Ref.~\onlinecite{norman2007} for the damping $\gamma$ and the mass renormalization constant $\lambda$. 
For $\omega$ larger than the energy of the fluctuations coupled to the electrons (phonons, density fluctuations) $\lambda\sim 0$, $\gamma$ becomes frequency independent, and its temperature dependence is
\begin{equation}
\gamma(T) = 2\gamma\left[1-\frac{2k_BT}{\omega_2}\ln\left(1-e^{-\omega_1/k_BT}\right)\right]
\label{normaneq}
\end{equation}
The parameter values relevant for the present materials are $\epsilon_\infty=4$, $\gamma=0.25$~eV, $\omega_p=3.1$~eV, $\omega_1=15$~meV and $\omega_2=300$~meV. 
The integration of the corresponding loss function is shown in Fig.~\ref{norman} where the temperature dependence was extracted by a power-law fit yielding an exponent of $2$.
\begin{figure}[h!]
\begin{center}
\includegraphics[width=0.8\columnwidth]{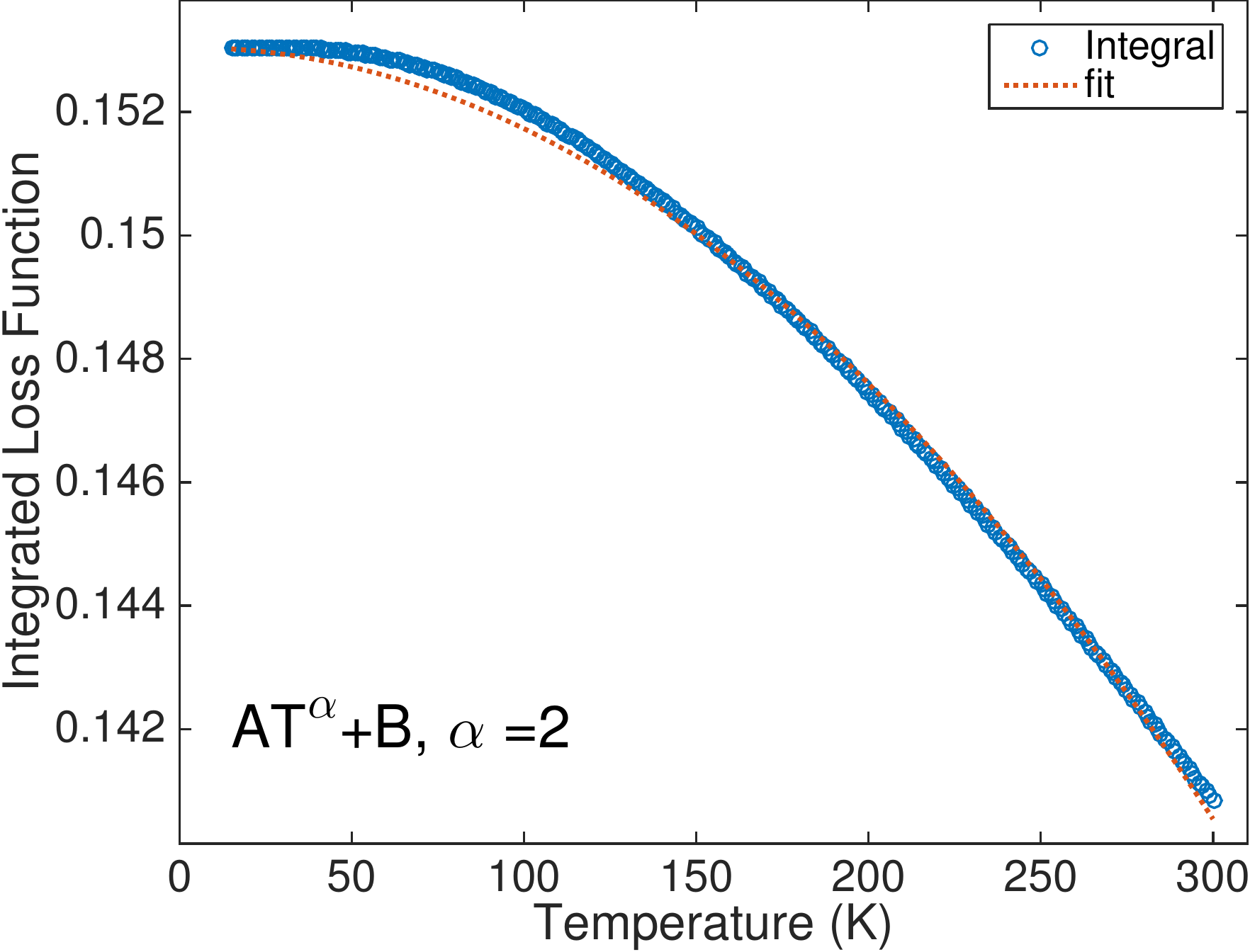}
\caption{Partial integral of the loss function as a function of temperature obtained by using the scattering rate in a Drude model of Norman~\emph{et al.}~\cite{norman2007}.}
\label{norman}
\end{center}
\end{figure}
\bibliographystyle{apsrev4-1}
\end{document}